\documentclass{aa}

\usepackage{txfonts}

\usepackage{graphicx}

\usepackage{natbib}
\bibpunct{(}{)}{;}{a}{}{,} 

\newcommand{\teff}{$T_\mathrm{eff}$}
\newcommand{\logg}{$\log g$}

\newcommand{\porb}{$P_\mathrm{orb}$}

\newcommand{\phiorb}{$\phi_{98}$}
\newcommand{\phirot}{$\phi_{67}$}

\newcommand{\msun}{M$_\odot$}
\newcommand{\rsun}{R$_\odot$}

\newcommand{\chisq}{$\chi^2$}
\newcommand{\chisqr}{$\chi^2_\nu$}
\newcommand{\china}{$\chi^2_\mathrm{Na}$}
\newcommand{\chica}{$\chi^2_\mathrm{Ca}$}

\newcommand{\oxygen}{$\mathrm{O}_2$} 
\newcommand{\water}{$\mathrm{H}_2\mathrm{O}$}

\newcommand{\hbeta}{\mbox{H$\beta$}}

\newcommand{\oex}{\mbox{\object{EX\,Hya}}}
\newcommand{\oexe}{\mbox{\object{EX\,Hydrae}}}

\newcommand{\kms}{km\,s$^{-1}$}

\newcommand{\ergs}{erg\,cm$^{-2}$s$^{-1}$}
\newcommand{\ergsa}{erg\,cm$^{-2}$s$^{-1}$\AA$^{-1}$}

\begin{document}

\title{\mbox{High-resolution spectroscopy of the intermediate polar EX Hydrae}:
I. Kinematic study and Roche tomography\thanks{Based on observations
collected with the ESO Very Large Telescope, Paranal, Chile, in
program 072.D--0621(A).}}

\author{K.~Beuermann\inst{1} \and K.~Reinsch\inst{1}}
  
%  offprint comment removed because we do not distribute reprints anymore !
%  \offprints{beuermann@astro.physik.uni-goettingen.de}
  
\institute{Institut f\"ur Astrophysik, Friedrich-Hund-Platz~1, D-37077~G\"ottingen, Germany,
\email{beuermann@astro.physik.uni-goettingen.de reinsch@astro.physik.uni-goettingen.de}}

\date{Received November 7, 2007\ / Accepted December 21, 2007}
  
% Abstract: Context, Aims, Methods, Results, [Conclusions]
\abstract { % Context 
\oex\ is one of the few double-lined eclipsing cataclysmic variables
that allow an accurate measurement of the binary masses.}
{ % Aims 
We analyze orbital phase-resolved UVES/VLT high resolution
($\lambda/\Delta \lambda \simeq 27000$) spectroscopic observations of
\oex\ with the aims of deriving the binary masses and obtaining a
tomographic image of the illuminated secondary star.}
{ % Methods
We present a novel method for determining the binary parameters by
directly fitting an emission model of the illuminated secondary star
to the phase-resolved line profiles of NaI$\lambda8183/8195$ in absorption
and emission and CaII$\lambda8498$ in emission.}
{ % Results 
The fit to the NaI and CaII line profiles, combined with the published
$K_1$, yields a white-dwarf mass \mbox{$M_1=0.790\pm0.026$\,\msun}, a
secondary mass $M_2=0.108\pm0.008$\,\msun, and a velocity amplitude of
the secondary star $K_2=432.4\pm4.8$\,\kms. The secondary is of
spectral type dM$5.5\pm 0.5$ and has an absolute $K$-band magnitude of
$M_\mathrm{K}=8.8$. Its Roche radius places it on or very close to the
main sequence of low-mass stars. It differs from a main sequence star
by its illuminated hemisphere that faces the white dwarf. The
secondary star contributes only 5\% to the observed spin-phase
averaged flux at 7500\AA, 7.5\% at 8200\AA, and 37\% in the
$K$-band. We present images of the secondary star in the light of the
NaI doublet and the CaII emission line derived with a simplified version of
Roche tomography. Line emission is restricted to the illuminated part
of the star, but its distribution differs from that of the incident
energy flux.}
{ % Conclusions 
We have discovered narrow spectral lines from the secondary star in
\oex\ that delineate its orbital motion and allow us to derive
accurate masses of both components. The primary mass significantly
exceeds recently published values. The secondary is a low-mass main
sequence star that displays a rich emission line spectrum on its
illuminated side, but lacks chromospheric emission on its dark side.}
   
\titlerunning{Kinematic study of the cataclysmic variable EX Hydrae}
\authorrunning{K.~Beuermann \& K.~Reinsch}

\maketitle

% ===========================================================================

\section{Introduction}

\oexe\ (orbital period \porb~=~98\,min) is the prototype of the short
period version of intermediate polars, the subclass of cataclysmic
variables in which a magnetic white dwarf accretes from a surrounding
gaseous disk or ring. Absorption or emission lines from the secondary
star have not been convincingly detected so far, because of the strong
veiling from the accretion disk and magnetic funnel
\citep{dhillonetal97,eisenbartetal02,vandeputteetal03}.  Consequently,
reports on \mbox{the masses of the} components have been controversial
\citep{hellieretal87,fujimotoishida97,hurwitzetal97,allanetal98,cropperetal98,
cropperetal99,belleetal03,beuermannetal03,hoogerwerfetal04}.

In this paper, we report high-resolution phase-resolved blue and red
spectrophotometry of EX Hya that reveals narrow lines from the
secondary star, notably KI$\lambda7665/7699$ and NaI$\lambda8183/8195$
in absorption and emission, a narrow emission component of
CaII$\lambda8498$, and a forest of faint emission lines from neutral
and singly ionized metals. These lines combine to define a unique
value for the radial velocity amplitude $K_2$ of the secondary and,
combined with the published $K_1$, allow us to derive accurate masses
of both binary components. Our approach involves a Roche tomographic
analysis of the illuminated secondary star and the synthesis of the
complex line profiles. The primary mass of 0.79\,\msun\ implies a
mass transfer rate close to that expected from gravitational
radiation.

The broad emission lines from the accretion disk and funnel that
dominate the blue spectra will be discussed elsewhere.

\section{Observations}
\label{sec:obs}
% -----------------------------------------------------------------------------
% Table 1
\begin{table}[b]
\caption{Journal of the VLT/UVES observations of EX Hya. Red and blue
spectra were acquired simultaneously (see text).}
\label{tab:obslog}
\begin{tabular}{llccc} 
\hline \hline \noalign{\smallskip}
Date & Target & UT & Number of & Exposure\\ 
     &        &    & spectra   & (s)\\ 
\noalign{\smallskip} \hline
\noalign{\smallskip}
Jan 23, 2004 & EX Hya & 5:42 -- 8:17 & 26 & 300 \\[0.5ex]
Jan 26, 2004 & EX Hya & 6:25 -- 8:38 & 22 & 300 \\[0.5ex]
Jan 29, 2004 & Gl300  & 11:09 --11:25 &  4 & 60/300 \\[0.5ex]
\noalign{\smallskip} \hline      
			         
\end{tabular}
\end{table}
% -----------------------------------------------------------------------------
% Fig. 1
\begin{figure*}[t]
\vspace{0.4mm}
\includegraphics[width=9.43cm]{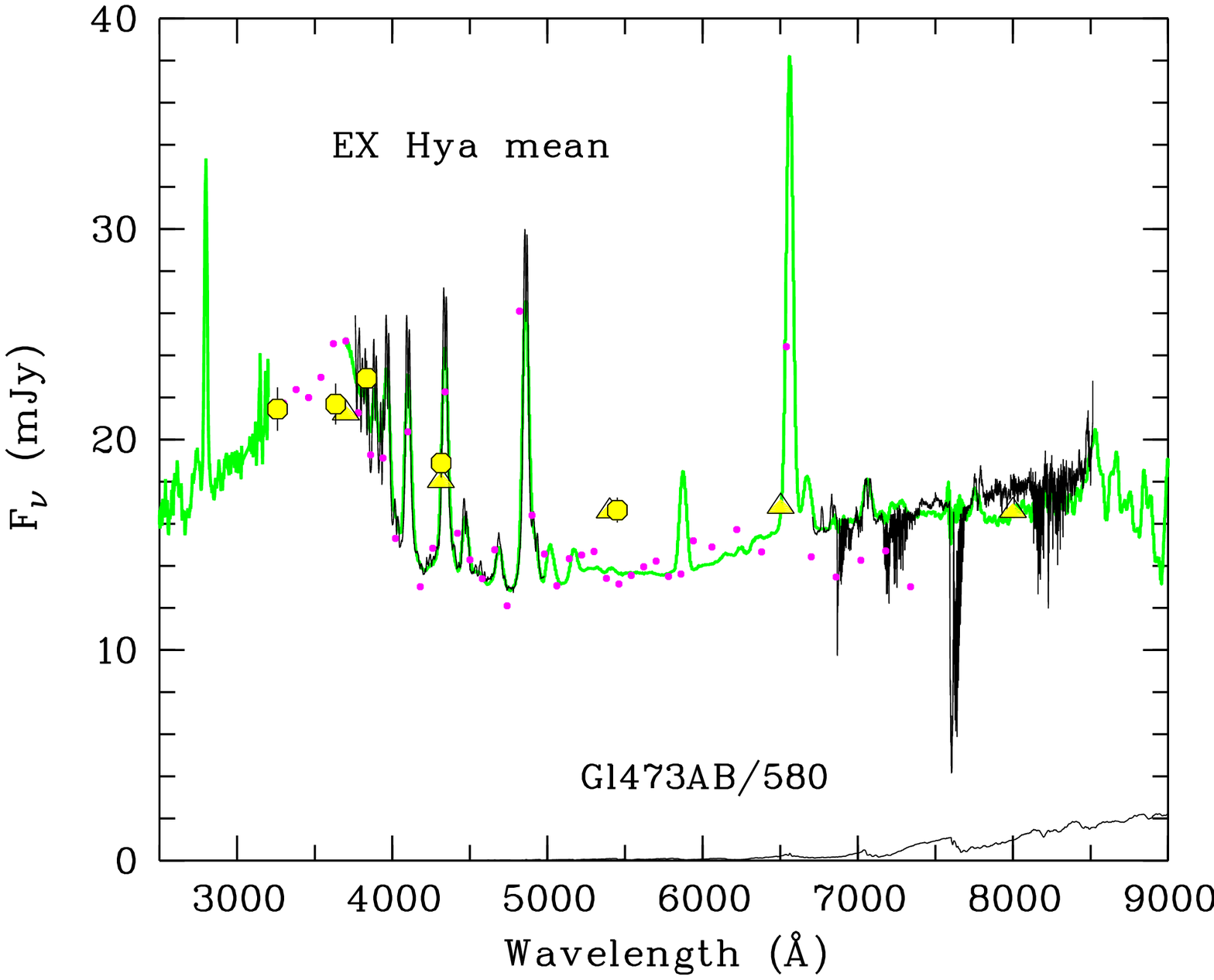}
\hspace*{-1.0mm}
\includegraphics[width=8.312cm]{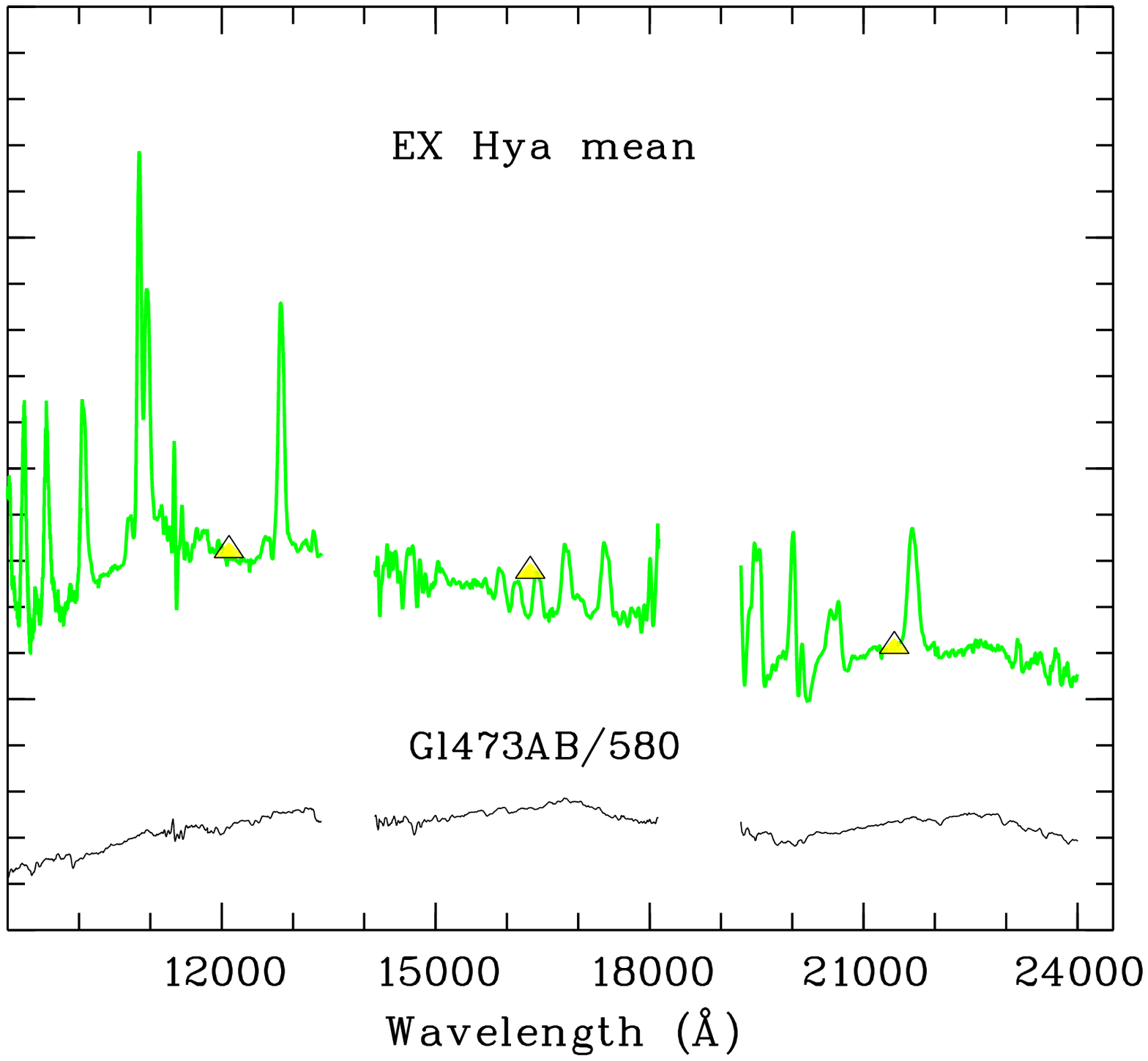}
\caption{Flux calibrated spin-averaged blue and red UVES spectra of EX
Hya (black curves) superposed on the overall spectral energy
distribution of \citet{eisenbartetal02} (green curves). Also
shown is the spectrophotometry of \citet{bathetal81} (small dots, see
text), the Walraven photometry of \citep{siegeletal89} (open circles),
and unpublished UBVRIJHK photometry (open triangles, see text). Also
shown is the optical/infrared spectrum of the dM5.5 star Gl473AB
\citep{eisenbartetal02} adjusted to the flux level of the secondary
star in EX Hya.}
\label{fig:meanspec}
\end{figure*}

EX Hya was observed in service mode with the UVES spectrograph at the
Kueyen (UT2) unit of the ESO Very Large Telescope, Paranal/Chile, on
January 23 and 26, 2004. Table 1 contains a log of the
observations. We adopted the ESO UVES pipeline reduction that provides
flat fielding, performs a number of standard corrections, extracts the
individual echelle orders, and collects them into a combined
spectrum. The flux calibration, except for the seeing correction
(Sect.~\ref{sec:dimm}), is also part of the pipeline reduction. The
wavelength calibration is derived from spectra taken with a ThAr
lamp. Blue and red spectra were measured simultaneously in the
wavelength ranges 3760--4980\AA\ and 6706--8521\AA\ with pixel sizes
of 0.030 and 0.041\AA, respectively.  With a slit width of 1\arcsec,
the FWHM resolution is 0.175\AA\ and the spectral resolution is
$\lambda/\Delta \lambda \simeq 47\,000$. To improve the S/N ratio, we
rebinned the data into 0.3\AA\ bins ($\sim10$\,\kms) and obtained an
effective resolution $\lambda/\Delta \lambda \simeq 27\,000$. Exposure
times were 300\,s with dead times of typically 60\,s between exposures
due to readout. The resulting orbital phase resolution is $\Delta
\phi_{98}=0.061$, the spin phase resolution is $\Delta
\phi_{67}=0.090$. For the subsequent analysis, we combined the data of both
nights into 16 orbital phase bins.

Figure~\ref{fig:meanspec} (left panel) shows the flux calibrated and
spin-averaged blue and red UVES spectra of EX Hya before the
correction for telluric-line absorption. The spectra are overlaid on
the spin-averaged spectral energy distribution from
\citet{eisenbartetal02} and the spectrophotometry of
\citet{bathetal81}. The latter is converted from spin phase
\phirot=0.15 to 0.25 (i.e., to spin average) using the known
wavelength dependence of the spin modulation. Also shown is the
Walraven photometry of \citet{siegeletal89} and the spin averages of
unpublished phase-resolved simultaneous UBVRIJHK photometry taken in
1982 by Joachim Krautter and Nikolaus Vogt (private
communication). The flux level of the mean red UVES spectrum agrees
with the earlier data to better than 10\%, whereas a
wavelength-dependent correction between -5\% and 20\% was needed to
adjust the mean blue UVES spectrum to the Eisenbart et al. and the
Bath et al. spectrophotometry. The high degree of internal consistency
between the various measurements provides confidence in the flux
calibration of the red UVES spectra discussed in this paper.

\begin{figure}[t]
\includegraphics[width=8.24cm]{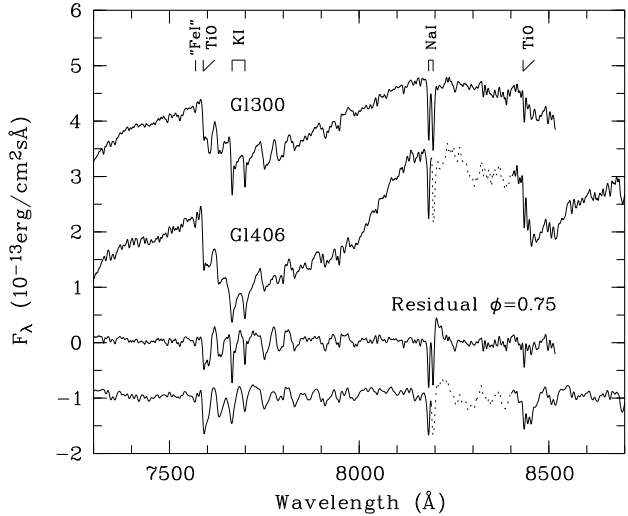}
\caption{Flux calibrated red UVES spectra of the M6 dwarf Gl406 and
the M425 dwarf Gl300 (shifted upwards by two units). The 'residual'
spectra shown as the two bottom curves refer to our method of
analyzing the EX Hya spectra and are explained in Sect.~\ref{sec:method}.  
The bottom one refers to Gl406 and is shifted downwards by one unit.}
\label{fig:mstars}
\end{figure}

Spectra of the M4.25 dwarf Gl300 were taken with the same setup on
January 29. A UVES spectrum of the M6 dwarf Gl406, more akin to the
secondary star in EX Hya than Gl300, was kindly provided by Ansgar
Reiners. Flux calibrated red spectra of these M-stars (corrected for
telluric lines) are displayed in Fig.~\ref{fig:mstars}. The gap in the
Gl406 UVES spectrum between 8190 and 8403\AA\ is filled in with a
medium-resolution archival spectrum of ours (dotted curve). The red
spectrum of Gl300 fits the Kron-Cousins $I_\mathrm{c}$-band photometry
($I_\mathrm{c}=9.17$, Leggett 1992) without any adjustment confirming
the accuracy of the UVES flux calibration, whereas the blue spectrum
shows a similar moderate loss of blue light as noted above. The Gl406
spectrum is adjusted to $I_\mathrm{c}=9.39$ (Leggett 1992). As
discussed below, the secondary star in EX Hya is of spectral type dM5
to dM6 and contributes about 5\% of the mean flux of EX Hya at
7500\AA. The correspondingly adjusted spectrum of the dM5.5 star
Gl473AB \citep{eisenbartetal02} illustrates the dominance of the disk
and funnel emission in EX Hya (Fig.~\ref{fig:meanspec}).

% Fig. 2

\section{Data Analysis}
\label{sec:dataanalysis}

In this section, we discuss the accuracy of the wavelength
calibration, outline the seeing correction that is part of the flux
calibration, and discuss the correction of the individual red spectra
for telluric line absorption.

\subsection{Wavelength calibration}
\label{sec:waca}

\begin{figure}[t]
\vspace*{1mm}
\includegraphics[width=8.8cm]{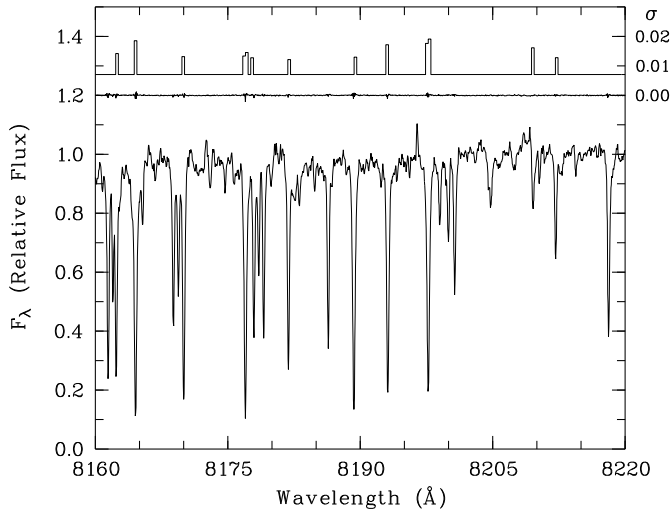}
\caption{Mean normalized spectrum of EX Hya in the vicinity of the
NaI$\lambda8183/8195$ doublet. Wavelength bins are 41\,m\AA\ and the
FWHM spectral resolution is 175\,m\AA. The two spectra shown at the
top are explained in Sect.~\ref{sec:method}. The NaI doublet becomes
detectable only after telluric correction.}
\label{fig:telluric}
\end{figure}

The UVES pipeline reduction provides wavelength-calibrated
spectra. The accuracy of this wavelength calibration can be verified
in the red spectral region by measuring the positions of the numerous
telluric lines of \oxygen\ and \water. As an example,
Fig.~\ref{fig:telluric} shows the water vapor lines near
8200\AA. Comparison of theoretical and observed positions of unblended
lines with moderate optical depths yield a mean difference in these
positions of $\Delta \lambda = 3\pm3$\,m\AA\ throughout the red
spectral range and a standard deviation of $\sigma = 13$\,m\AA. Hence,
the wavelength calibration is quite accurate and the systematic error
in the derived radial velocities does not exceed 1.0\,\kms.

\subsection{Seeing correction}
\label{sec:dimm}

\begin{figure}[t]
\vspace{0.4mm}
\includegraphics[width=8.7cm]{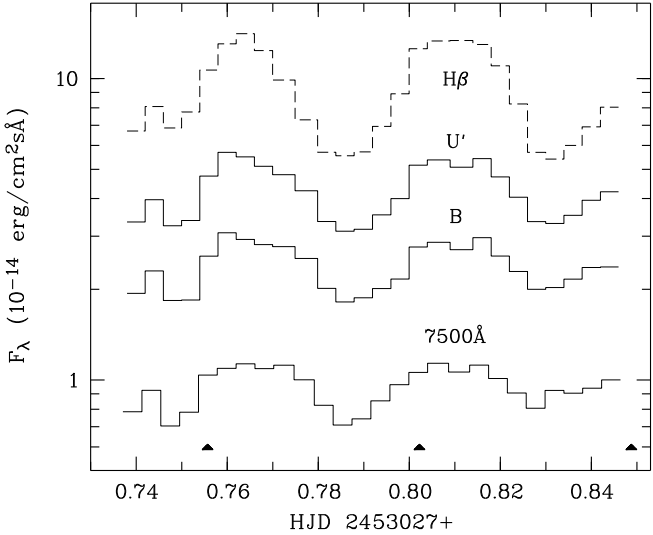}
\caption{Flux-calibrated light curves of EX Hya obtained on January
23, 2004. The bands $U'$ and $B$ refer to the mean flux in the
intervals 3770--4000\AA\ and 4000--4800\AA, respectively. The
integrated flux of the \hbeta\ emission line is given in units of
$10^{-13}$\,\ergs. The triangles indicate the times of spin maxima
predicted by the quadratic ephemeris of \citet{helliersproats92}.}
\label{fig:lc}
\end{figure}

Seeing information is provided by the DIMM on Paraneal
\citep[Difference Image Motion Monitor,][]{sarazinroddier90} and by
the width of the individual spectral images perpendicular to the
dispersion direction. Both methods yielded compatible estimates of the
FWHM seeing that varied between 0.9\arcsec\ and 2.2\arcsec\ on January
23 and between 0.7\arcsec\ and 1.2\arcsec\ on January 26.  We
corrected each spectrum for the seeing losses assuming a Gaussian
point spread function and a central position of the source in the 1.0
arcsec slit. This correction is part of our flux calibration and is
included whenever we quote absolute fluxes. In particular, this holds
for the spectra in Fig.~\ref{fig:meanspec} and for the light curves
shown in Fig.~\ref{fig:lc} and discussed in Sect.~\ref{sec:ephemeris},
below.

% Fig. 3

\subsection{Telluric line correction}
\label{sec:method}

We reconstruct the stellar flux incident on the atmosphere by
correcting the observed spectra for the absorption in the telluric
lines. Given the structure of the spectrum as shown in
Fig.~\ref{fig:telluric}, this is a formidable task, in particular,
since the water vapor content of the atmosphere on January 23 and 26,
2004, was rather high. Absorption in the strongest \water\ lines near
8200\AA\ reached 88\%, whereas the depth of the NaI$\lambda8183/8195$
doublet is only 2.6\% of the continuum.

Telluric correction of the observed spectrum $F_\mathrm{i}(\lambda )$
of orbital phase bin $i$ requires division by
$\mathrm{exp}({-\tau_\mathrm{i}(\lambda)}$), where
$\tau_\mathrm{i}(\lambda)$ is the wavelength dependent optical depth
of the atmosphere. We need to define a template $\tau(\lambda)$ that
allows us to construct the individual $\tau_\mathrm{i}(\lambda)$ and
to reach the desired accuracy of the reconstructed fluxes.
To start with, we employed the normalized spectrum of a featureless
standard star obtained in a different night as template
$\tau(\lambda)$. This proved unsuccessful, because the optical depth
in water vapor was only about 1/3 of that in our observations and the
optical depth structures did not match. We resorted, therefore, to 
an entirely different method using the orbital mean red spectrum of
each night as a telluric template.  At first glance, this might seem
inappropriate, because it removes all spectral structure from the mean
of our set of phase-resolved spectra. For the special case of a
cataclysmic variable with its rapid variability, the method is
attractive, however, because it removes all spectral structure that
does not display orbital (or more generally temporal)
variability. Although the mean spectrum vanishes, the individual
corrected exposures that we refer to as 'residual' spectra retain much
of the spectral structure at wavelength scales less than that of the
radial velocity variation of $\sim 20$\AA\ in case of EX Hya. 

To explain the method, we present a model calculation with the~4-\AA\
binned spectra of Gl300 and Gl406 shown in Fig.~\ref{fig:mstars}. Let
$F(\lambda )$ be the spectrum of the M-star and
$F_\mathrm{i}(\lambda)$ the same spectrum for orbital phase bin
$i$ shifted in wavelength according to the radial velocity of the
secondary star in EX Hya. The mean of $N$ phase bins is
$\overline{F}(\lambda)=(1/N)\sum_1^{N}F_\mathrm{i}(\lambda )$ and the
residual spectra are
$f_\mathrm{i}(\lambda)=F_\mathrm{i}(\lambda)-\overline{F}(\lambda )$.
Figure~\ref{fig:mstars} (bottom curves) shows the residuals of Gl300
and Gl406 at orbital phase $\phi_\mathrm{98,\,i}=\,0.75$, calculated
for an orbital velocity amplitude of 432.4\,\kms\ and a spectral flux
that does not vary with phase. Spectral structure that extends over
$\Delta \lambda>20$\AA\ is lost, while structure on a shorter scale is
preserved with an amplitude up to 90\% of the original.

% Fig. 4

\begin{figure*}[t]
\vspace{12mm}
\hspace{11.73mm} 
\includegraphics[width=13.6mm,clip]{9010f5a.ps}
\hspace{1.6mm}
\includegraphics[width=150.8mm,clip]{9010f5b.ps}

\vspace{-59.6mm}
\includegraphics[width=26.8mm]{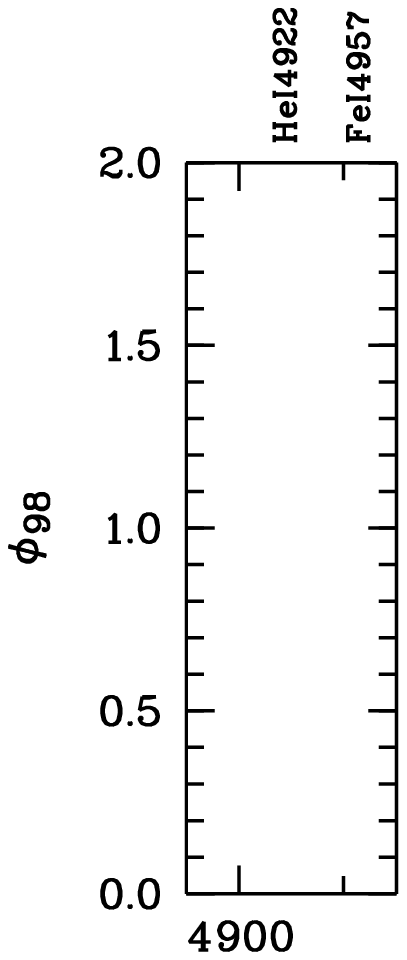}
\hspace{0.8mm}
\includegraphics[width=151.3mm]{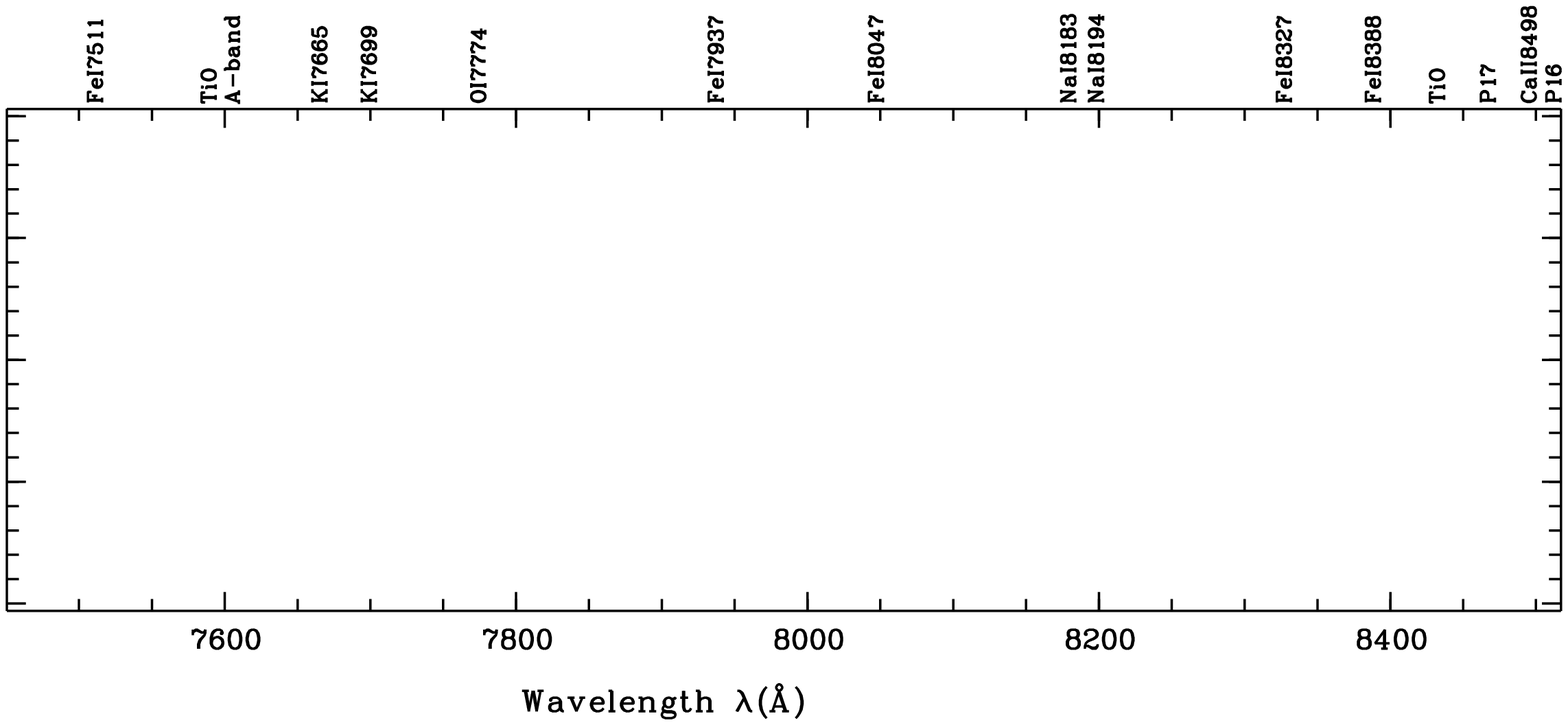}
\caption{Phase-resolved and telluric line-corrected 'residual' spectra
of EX Hya collected into 16 phase bins and repeated for two orbits
(see Sect.~\ref{sec:method}). Numerous emission and absorption
features from the secondary star are visible, most prominent the TiO
band heads at 7589 and 8432\AA, the KI$\lambda7665/7699$ and
NaI$\lambda8183/8195$ emission/absorption lines, and CaII$\lambda
8498$ in emission. The broad HeI, OI, and Paschen lines are from the
accretion disk and funnel. The displayed intensities are relative to
the orbital mean, the zero level is shaded gray, lower and higher
intensities appear darker and lighter, respectively.}
\label{fig:phased}
\end{figure*}

For EX Hya, we express the 'raw' spectrum $F_\mathrm{i}(\lambda)$ of
orbital phase bin $i$ as
\begin{equation}
F_\mathrm{i}(\lambda) = C_\mathrm{i}(\lambda)\,n_\mathrm{i}(\lambda),
\end{equation}
where $C_\mathrm{i}(\lambda)$ represents the smooth wavelength
\mbox{dependence of} the continuum and $n_\mathrm{i}(\lambda)$ is the
normalized spectrum that contains all spectral structure. We use the
orbital mean of the $n_\mathrm{i}(\lambda)$ as the telluric template $
\overline{n}(\lambda)\equiv \mathrm{exp}(-\tau(\lambda))$. The
correction function for phase bin $i$ is calculated as
$\mathrm{exp}(-\tau_\mathrm{i}(\lambda))=\mathrm{exp}(-\alpha_\mathrm{i}\tau(\lambda))=\overline{n}(\lambda)^{\,\alpha_\mathrm{i}}$
with a parameter $\alpha_\mathrm{i}$ that is independent of
wavelength. We determine the $C_\mathrm{i}(\lambda)$ by fitting a
low-order polynomial to the continuum of the raw spectra.  The
template is then fitted to the resulting $n_\mathrm{i}(\lambda)$ over
a specified spectral range allowing for a small wavelength shift
$\delta \lambda_\mathrm{i}$ (of order m\AA). Hence, the residual
spectrum for phase bin $i$ is
\begin{equation}
f_\mathrm{i}(\lambda) =
F_\mathrm{i}(\lambda){[\,\overline{n}\,(\lambda+\delta
\lambda_\mathrm{i})]^{-\alpha_\mathrm{i}}}-C_\mathrm{i}(\lambda )~,
\label{eq:normspec}
\end{equation}
with fit parameters $\alpha_\mathrm{i}$ and $\delta
\lambda_\mathrm{i}$. In our data, the optical depth varies little with
orbital phase and the $\alpha_\mathrm{i}$ stay close to unity. We
define the continua $C_\mathrm{i}(\lambda)$ by quadratics fitted to
the $F_\mathrm{i}(\lambda)$ at 6800\AA, 7720\AA, and 8362\AA\ for
$\lambda<8000$\AA\ and at 7130\AA, 7854\AA, and 8362\AA\ for
$\lambda>8000$\AA. At these wavelengths, any temporal variation of the
spectral flux is removed and the residual spectra equal zero. Any
remaining orbital modulation of the residual flux is relative to these
wavelengths.
Telluric absorption is considered separately for absorption by
\oxygen\ and \water\ lines. The template is fitted to the \oxygen\
lines between 7620\AA\ and 7657\AA\ avoiding the KI lines and the
deepest part of the A-band at 7593--7617\AA. It is fitted to the
\water\ lines over the bands 8000-8150\AA\ and 8220--8350\AA\ avoiding
the NaI doublet. We restrict the subsequent analysis to
$\lambda>7300$\AA\ and piece the residual spectra together from the
\oxygen-corrected part below 8000\AA\ and the \water-corrected part
above 8000\AA. The two fits remove the weaker telluric lines outside
the immediate fit intervals almost perfectly and only the
7593--7617\AA\ section remains problematic, where absorption at the
line centers reaches 99.8\%. The spectra of both nights are then
combined and collected into 16 phase bins.  The result is shown in
Fig.~\ref{fig:phased}. Spectral structure on a short wavelength scale
is preserved in the individual reconstructed spectra although they
average to zero.

\subsection{Statistical noise}

A natural consequence of \emph{any} telluric-line correction is the
increased noise level of the restored fluxes at the positions of the
strong lines. We opted to keep all data points in the resulting
spectra and account for the enhanced noise at the telluric line
positions by the increased statistical errors of the affected data
points.  The orbital mean of the reconstructed normalized spectra for
the first night is shown in Fig.~\ref{fig:telluric} (second curve from
top, shifted upwards by 0.2 units). As expected, it equals unity with
only minute deviations that disappear, too, when it is rebinned to
0.3\AA. The statistical error of the individual 0.3\AA\ spectral bins
is determined from the mean rms noise in regions that avoid the
strongest telluric lines and amounts to 0.7\% of the continuum
(Fig.~\ref{fig:telluric}, uppermost curve). It reaches 2\% in the
deepest \water\ lines around 8200\AA\ and huge values in the deepest
lines of the A-band between 6593 and 7617\AA. The 0.7\% level
corresponds to an absolute error of $6\times 10^{-17}$\ergsa\ near
8200\AA.

% Fig. 5

% Fig. 6
\begin{figure*}[t]
\vspace{0.4mm}
\includegraphics[width=8.4cm]{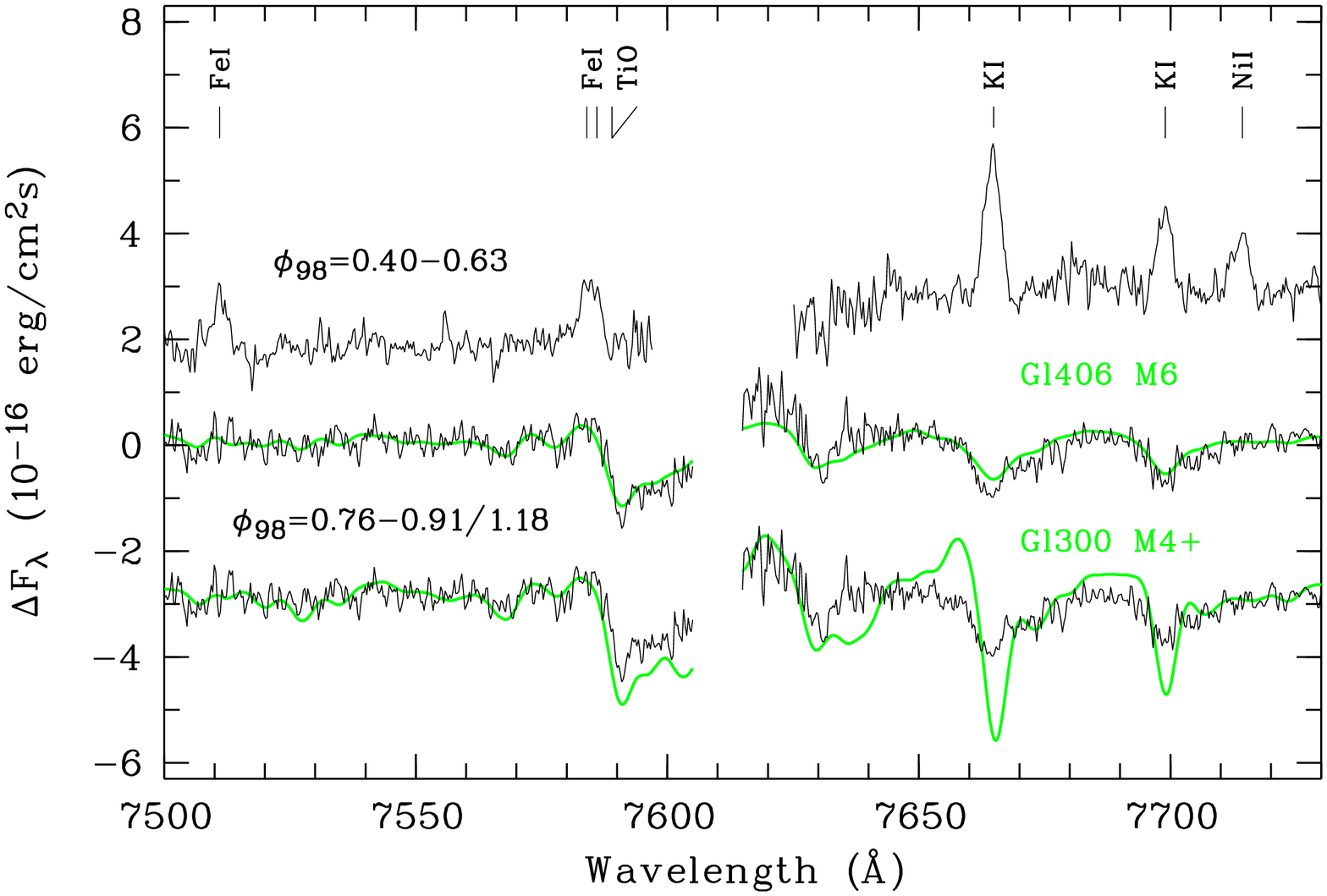}
\hspace*{1mm}
\includegraphics[width=4.2cm]{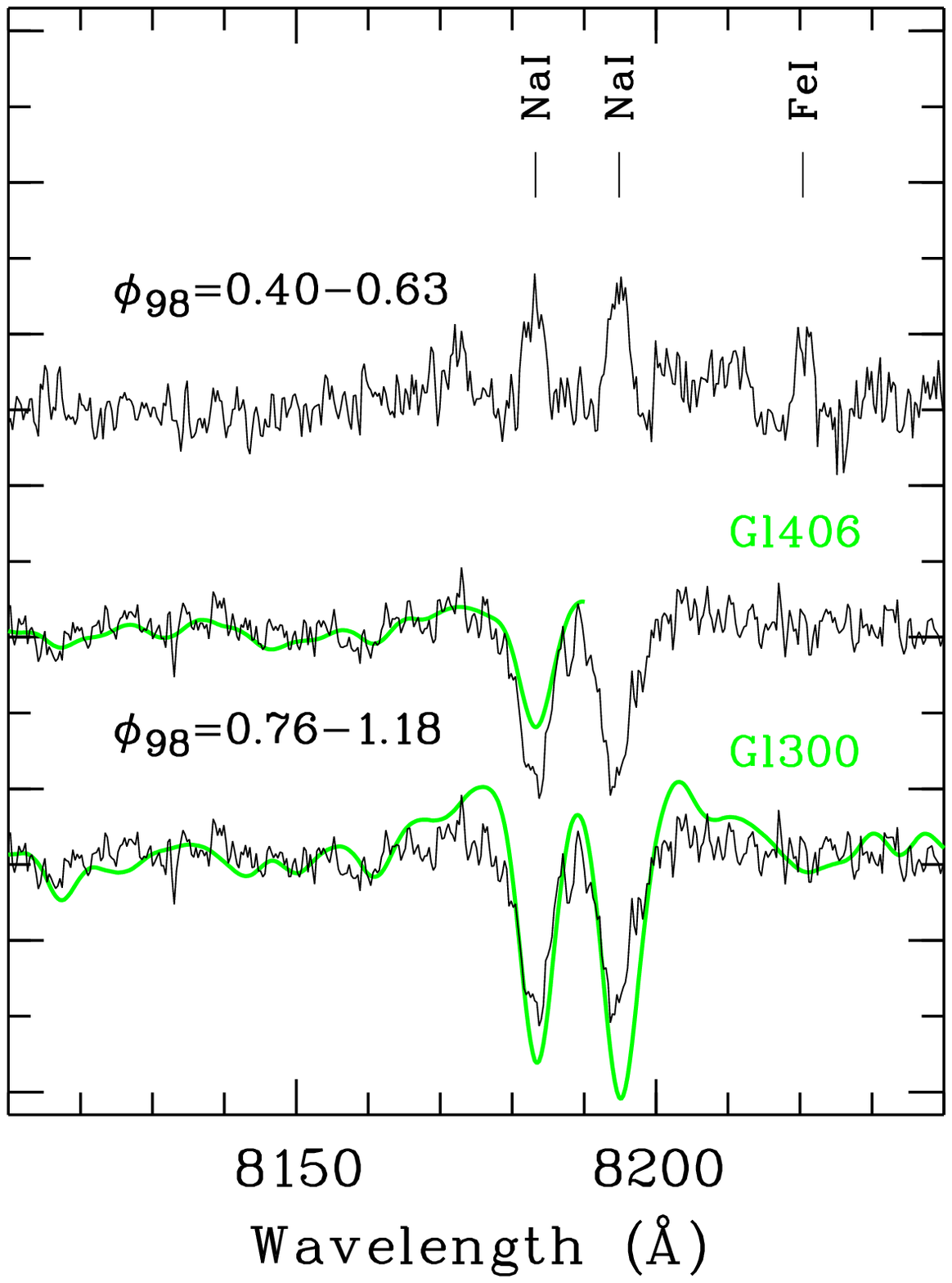}
\hspace*{1mm}
\includegraphics[width=4.85cm]{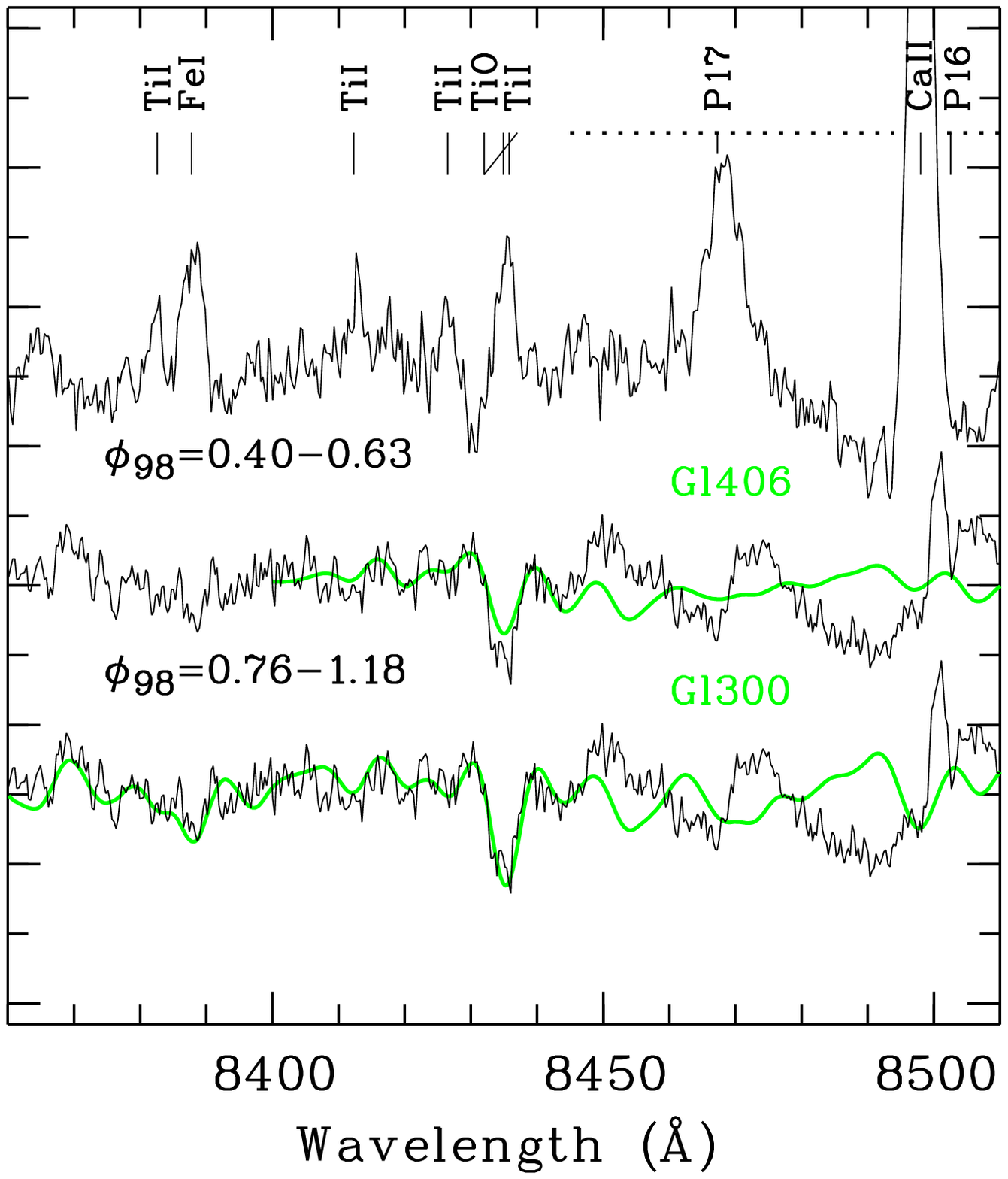}
\caption{Absolute 'residual' spectral fluxes $f_\mathrm{i}(\lambda)$
of EX Hya for the illuminated face of the secondary star (top, black
data train, shifted upwards by three units) and its dark side (center
and bottom, black data train, unshifted and shifted downwards by three
units, respectively). The spectra are shifted in wavelength to the
respective rest frame before averaging them over the orbital phase
intervals indicated in the figure (see text). Shown for comparison are
the correspondingly constructed 'residual' fluxes of the M6 dwarf
Gl406 and the M4.25 dwarf Gl300.}
\label{fig:secgliese}
\end{figure*}

\section{Observational results}
\label{sec:results}

The gray plots of Fig.~\ref{fig:phased} show the 16 phased-resolved
residual spectra for the wavelength intervals 4875--4975\AA\ and
7300--8520\AA\ displayed twice. The wavelength bins are 0.3\AA.  In
the selected spectral ranges, the strongest broad emission lines from
the accretion disk and funnel are HeI$\lambda4922$,
OI$\lambda7772/7774/7775$, Paschen P17, CaII$\lambda8498$, and Paschen
P16. The narrow spectral features from the secondary star include the
TiO band heads at 7589\AA\ and 8432\AA, the KI$\lambda7665/7699$ and
the NaI$\lambda8183/8195$ lines in absorption and emission,
CaII$\lambda8498$ in emission, and numerous metal emission lines,
among the stronger ones FeI$\lambda4957$, FeI$\lambda8327$ and
FeI$\lambda8388$. The absorption lines are strong between
\phiorb\,=\,0.65 and 1.35, whereas the emission components reach peak
flux at \phiorb~$\simeq0.50$, both indications of an origin from the
illuminated secondary star. The narrow and strong CaII$\lambda8498$
emission line is superimposed on a complex background of the much
broader Paschen P16, P17, and CaII$\lambda8498$ disk lines. We have
separated the narrow line interactively from the background and
included an estimate of the uncertainty of this procedure in the
errors of the individual data points.

\subsection{The secondary star in EX Hya}
\label{sec:secondary}

Figure~\ref{fig:secgliese} displays mean spectra for the phases when
the illuminated face of the secondary star or its dark side are in
view (black curves). The spectra are summed over the respective phase
intervals, are placed on an absolute flux scale, and shifted into
the rest systems of the emission or the absorption lines,
respectively. To appreciate the weakness of the lines note that the depth of
the NaI doublet is 2.6\% of the continuum at 8200\AA\
(Fig.~\ref{fig:meanspec}). Information on the spectral type of the
secondary star in EX Hya can be obtained from a comparison of these
spectra with those of the M-stars Gl300 and Gl406 (green curves). We
have used the M-star spectra at their full spectral resolution,
adjusted their fluxes with the ratio of the solid angles $\Omega$,
\begin{equation}
F_\mathrm{sec} = F_\mathrm{M}\,\Omega_\mathrm{sec}/\Omega_\mathrm{M} =
F_\mathrm{M}(R_\mathrm{sec}/d)^2(d_\mathrm{M}/R_\mathrm{M})^2,
\label{eq:sec_M}
\end{equation}
formed the residuals as explained in Sect.~\ref{sec:method}, and
rebinned them to 0.3\AA. Here, $d_\mathrm{M}$ and $R_\mathrm{M}$ are
the distance and the stellar radius of the M-star
\citep{beuermannetal99}, whereas $d=(64.5\pm1.2)$\,pc
\citep{beuermannetal03} and
$R_\mathrm{sec}=(0.1516\pm0.0034)\,R_\odot$ (see
Sect.~\ref{sec:system}, below) denote the distance of EX Hya and the
mean radius of its secondary, respectively. The ratio of the solid
angles amounts to
$\Omega_\mathrm{sec}/\Omega_\mathrm{M}=0.00421\pm0.00036$ for Gl300
and $\Omega_\mathrm{sec}/\Omega_\mathrm{M}=0.00170\pm0.00015$ for
Gl406. The residual spectrum of the dark side of EX Hya is displayed
twice in the three panels of Fig.~\ref{fig:secgliese}, at the center
and the bottom. For $\lambda<7605$\AA, we have chosen the phase
interval \phiorb=0.76--0.91, when the emission lines have already
disappeared and the blueshift has moved the TiO band head at 7589\AA\
away from the telluric A-band. For $\lambda>7615$\AA, we sum over the
entire dark side of the secondary (\phiorb=0.76\,--\,1.18). Our
comparison indicates that the residual spectrum of Gl406 reproduces  the TiO
band head of EX Hya at 7589\AA\ perfectly. The KI lines of Gl406 fit
almost as well, whereas those of Gl300 utterly fail. The NaI doublet
suggests a spectral type between Gl300 and Gl406 and the TiO band
head at 8432\AA\ prefers Gl300. Many of the low-amplitude wiggles in
the EX Hya spectrum faithfully reproduce rotationally smoothed
absorption line structures of the two M-stars. This agreement is lost
for $\lambda>8445$\AA, where the broad disk lines veil the
spectral features of the secondary star. The TiO band head at
8432\AA\ is also affected by the two flanking TiI emission lines and,
assigning it a lower weight, we conclude that the EX Hya spectrum is
much better represented by Gl406 (dM6) than by Gl300 (dM4.25). The
implied spectral type of the secondary is slightly earlier than M6 and
we settle for dM$5.5\pm0.5$.
Interpolating between the adjusted fluxes of Gl300 and Gl406, we find
that the the secondary contributes $0.92\pm 0.28$\,mJy and $1.40\pm
0.48$\,mJy at 7500\AA\ and 8200\AA, respectively. This corresponds to
$(5.3\pm1.6)$\% and $(7.5\pm2.6)$\% of the spin-averaged flux of EX
Hya, respectively (see Fig.~\ref{fig:meanspec}).

% Fig. 7
\begin{figure}[t]
\includegraphics[width=8.6cm]{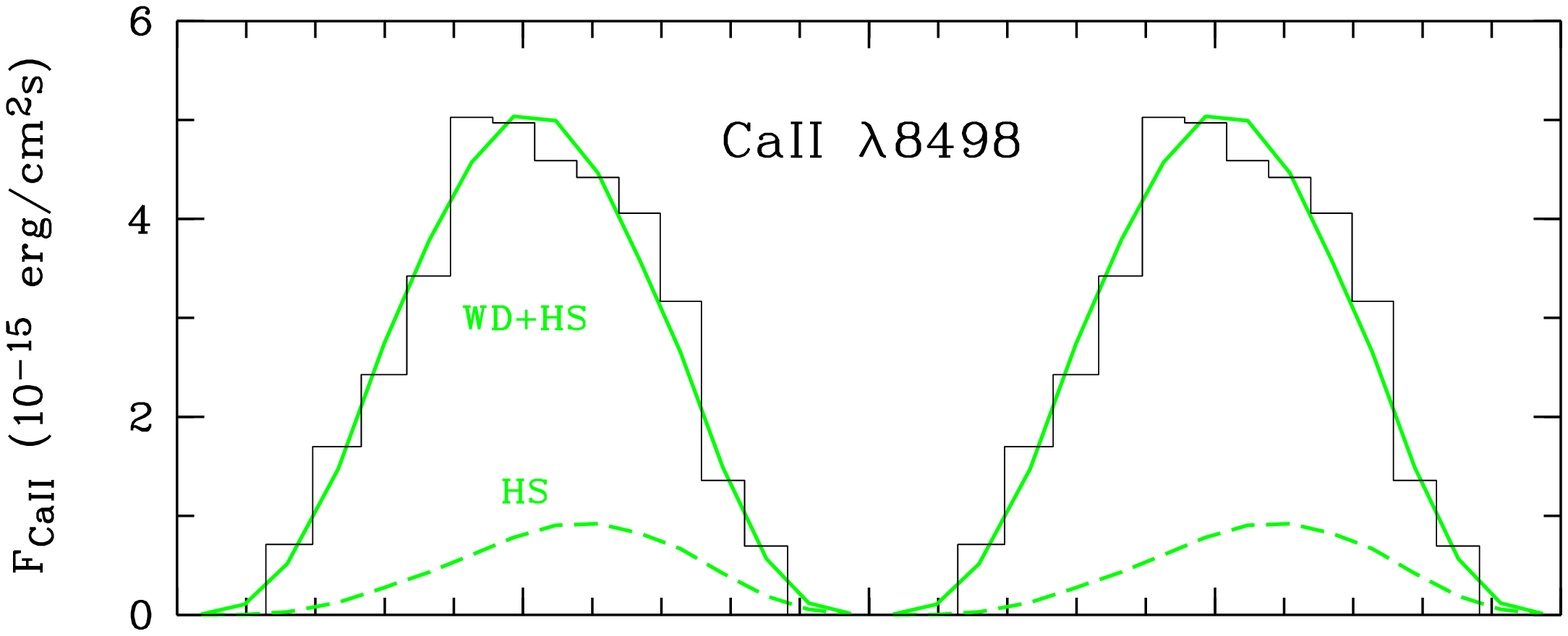}
\includegraphics[width=8.8cm]{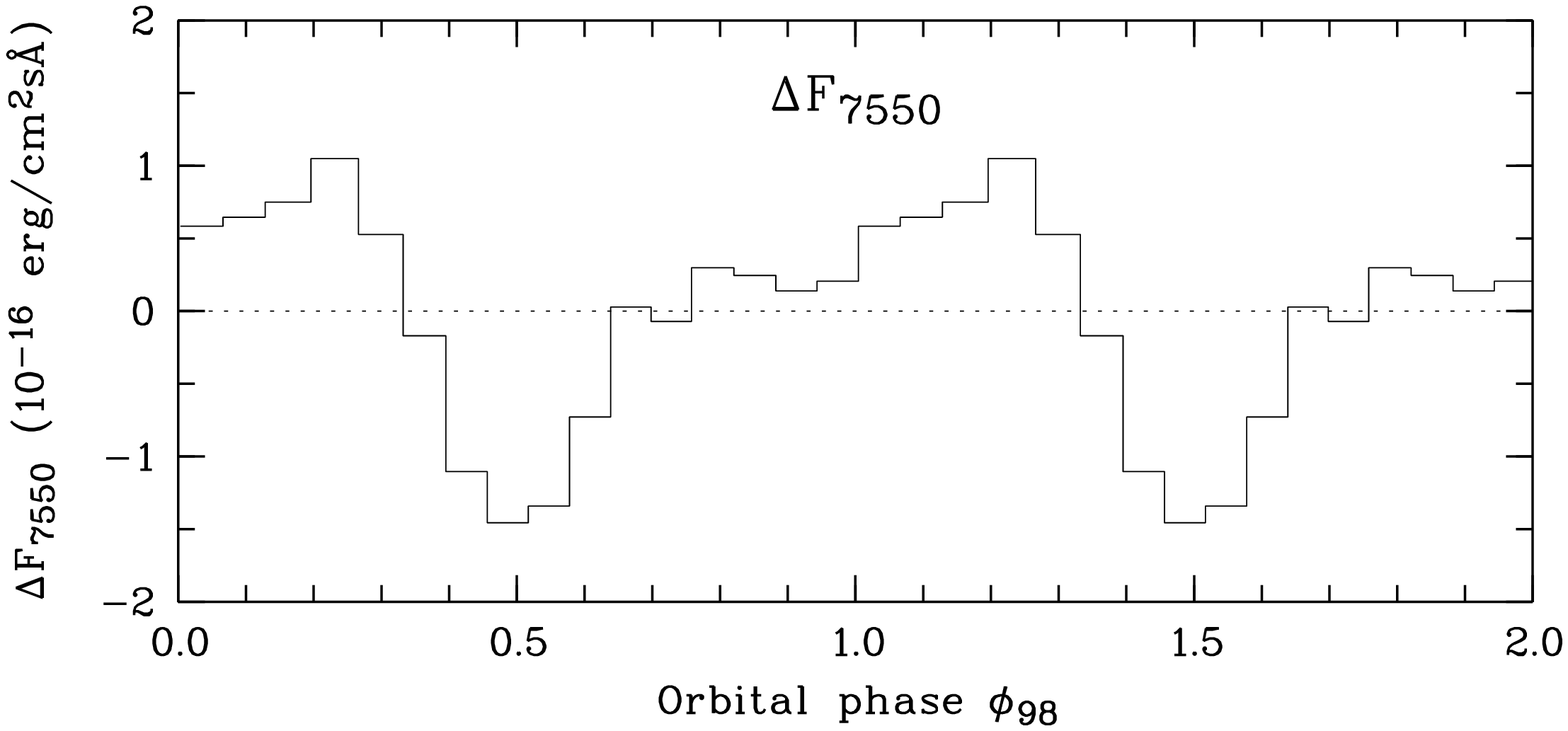}
\caption{\emph{Top: }Orbital light curve of the CaII$\lambda 8498$
integrated emission line flux (black histogram) and model fit (green
curves). The two components refer to the white dwarf (WD) and the hot
spot (HS) as irradiation sources. \emph{Bottom: }Flux at 7550\AA\
\emph{relative} to 7720\AA. The dip indicates a weakening of the
TiO$\lambda7589$ band strength on the illuminated face of the
secondary star. The abscissa is orbital phase calculated from
Eq.~\ref{eq:deltaphi}.}
\label{fig:calc}
\end{figure}

Although the dark side of the secondary looks like a dM5.5 star, this
does not hold for the illuminated side as shown by the spectrum for
\phiorb =\,0.40--0.63 in Fig.~\ref{fig:secgliese} (top). This phase
interval is characterized by the peak flux of the emission lines and a
disappearance of the TiO$\lambda7589$ absorption edge.
Figure~\ref{fig:calc} shows the light curves of the integrated flux of
the CaII$\lambda8498$ emission line (upper panel) and the difference
$\Delta F_{7550}$ of the residual fluxes at 7550 and 7720\AA\ (lower
panel). The minimum of the latter coincides with the emission line
maximum stressing the weakness of the TiO$\lambda7589$ band head on
the illuminated face, a result that is reminiscent of Wade \& Horne's
(1988) finding for Z~Cha.

Applying Eq.~(\ref{eq:sec_M}) to the $K$-band, allows us to estimate
the contribution of the secondary to the infrared flux of EX Hya.  The
apparent $K$-band magnitudes of Gl300 and Gl406, $K=6.68$ and 6.08,
imply $K_\mathrm{sec}=12.62$ and 13.00 for a secondary of spectral
type dM4.25 or M6, respectively. For the preferred spectral type
dM$5.5\pm0.5$, EX Hya B has $K_\mathrm{sec}=12.89\pm0.11$. We can
alternatively start from the surface brightness calibration of field
M-dwarfs by \citet{beuermann06} that implies
$S_\mathrm{K}=4.71\pm0.11$ for a spectral type dM$5.5\pm0.5$. With
$R_\mathrm{sec}=(0.1516\pm0.0034)\,R_\odot$ and $d=(64.5\pm 1.2)$\,pc,
we find
$K=S_\mathrm{K}-5\,\mathrm{log}[(R_\mathrm{sec}/R_\odot)(10\mathrm{pc}/d)]=12.86\pm0.13$,
nearly identical with the previous number. The absolute magnitude of
the secondary in EX Hya is $M_\mathrm{K}=8.83\pm0.12$, in perfect
agreement with that of a main sequence dM5.5 star.
The spin-averaged apparent $K$-band magnitude of EX Hya is
\mbox{$K=11.81$} (Fig.~\ref{fig:meanspec}) to which the secondary
contributes $(37.6\pm 4.2)$\%, somewhat less than estimated by
\citet{eisenbartetal02}. These authors noted already 
that part of the modulation of the infrared bands at one half the
orbital period that looks like the ellipsoidal modulation of the
secondary star may be produced by the bulge on the accretion disk. We
confirm this suspicion by noting that the optical bands of the
unpublished UBVRIJHK photometry (Sect.~\ref{sec:obs}) display a 49-min
modulation, too.

\subsection{Radial velocity amplitudes and systemic velocity}
\label{sec:vrad}

\begin{figure}[t]
\includegraphics[width=8.8cm]{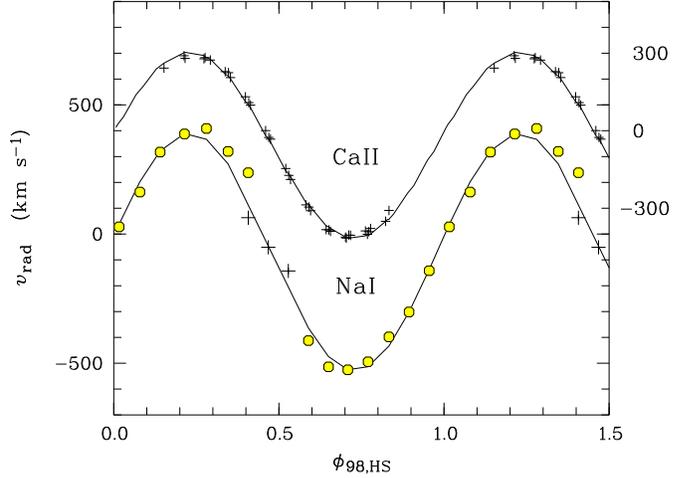}
\caption{Radial velocity curves of NaI$\lambda8183/8195$ in absorption
($\medcirc $) and emission ($+$) (left hand scale) and of the
CaII$\lambda8498$ emission line ($+$) (right hand scale) vs. orbital phase
$\varphi_\mathrm{98,HS}$ \citep{helliersproats92}. Solid lines
represent sinusoids least squares-fitted to the data (see
Sect.~\ref{sec:vrad}). The error bars are smaller than the data
points. }
\label{fig:vrad}
\end{figure}

The brightest emission lines suited for radial velocity measurements
are CaII$\lambda8498$, KI$\lambda7665/7699$, FeI$\lambda8327$, and
FeI$\lambda4958$. Of the absorption lines, only the
NaI$\lambda8183/8195$ doublet yields reliable results. In
KI$\lambda7665/7699$, the emission component is relatively stronger
and the profiles are affected by residuals of the telluric \oxygen\
lines. In this initial analysis, we measured radial velocities using
Gaussians for the single lines and a double Gaussian with a separation
of 11.57\AA\ for the NaI doublet. Figure~\ref{fig:vrad} shows the radial
velocity curves of the CaII emission line measured from the 48
individual exposures and of the NaI doublet measured from the 16
phase-binned spectra. The abscissa $\phi_\mathrm{98,HS}$ is the
orbital phase calculated from the linear ephemeris of
\citet{helliersproats92}. The NaI absorption component is disturbed by
emission on the descending branch and becomes undetectable around
$\phi_{98}\simeq0.5$. We have measured the velocity amplitudes, the
phase shifts, and the systemic velocities from sinusoidals fitted to
all data points for the emission lines and to the data points between
$\phi_\mathrm{98,HS}=0.71$ and 1.21 for NaI. The fit results are
listed in Tab.~\ref{tab:vrad}. Although these fits do not account for
the slight ellipticity visible in the data, they yield quite accurate
values of the velocity amplitudes $K_2'$ of the emission lines and of
the systemic velocity $\gamma$.  Neither the emission lines with an
average $K_2'=351$\,\kms\ nor the NaI absorption lines with $K_2''=460$\,\kms\
provide a reliable measure of the velocity amplitude $K_2$ of the
secondary star, which we expect to fall in between.
Note that the TiO$\lambda7589$ band head is not suited to track the
motion of the secondary star, because it is disturbed by the A-band
around \phiorb=0.25 and by FeI$\lambda7584/7586$ emission between
\phiorb=0.2 and 0.8.

% Fig. 8

The systemic velocities of the individual lines in Tab.~\ref{tab:vrad}
are generally in good agreement, although the NaI absorption line
value may be affected by the emission component and the first KI value
by possible blending with FeI at 7664.29\AA. Excluding these, the mean
systemic velocity of the remaining four lines becomes
\mbox{$\gamma=-59.0\pm 0.9$} \kms. The small systematic error of our
wavelength calibration adds only 0.5\,\kms\ to the error budget of
$\gamma$ (see Sect.~\ref{sec:waca}). Confirmation on the common origin
of all narrow emission lines and on the value of $\gamma$ is obtained
from the numerous lines seen in the blue spectra. They originate from
neutral or singly ionized species and are restricted to the phase
interval when the illuminated face of the secondary is in view. The
better defined ones with equivalent widths between 0.04\AA\ and 0.11\AA\
are HI$\lambda3770,3797,4102$,
FeI$\lambda3856,3889,4198,4271,4314,4426,4890,4957$, CaII\,K,
MnI$\lambda4034$, ArI$\lambda4045$, OII$\lambda4144$,
NaI$\lambda4216$, NeII/AlII$\lambda4227$, OI$\lambda4232$, NeI$\lambda
4274$, CrI$\lambda4583$, and MgII$\lambda4938$. Narrow emission
components are not detected in the HeI lines and in HeII$\lambda4886$.
We have studied these faint lines in a cursory manner shifting them
into their rest system with an appropriate $K_2'$. We obtain straight
vertical lines in the 2-D image for the $K_2'$ value of the red lines,
351\,\kms. The systemic velocities are measured from a spectrum
co-added over the phase interval of best visibility of the
lines. Their distribution has a standard deviation of $\sigma
=2.3$\,\kms\ and a mean $\gamma=-59.6$\,\kms, indistinguishable from
that of the red lines.

% -----------------------------------------------------------------------------
% Table 3
\begin{table}[t]
\caption{Results for sinusoidal fits
$\upsilon_\mathrm{rad}=K\,\mathrm{sin}[2\pi(\phi_{98}+\phi_0)]+\gamma$~~to
the radial velocities of narrow emission lines and the NaI absorption
line doublet. The errors in brackets refer to the
last digit. See text for systematic errors.}
\label{tab:vrad}
\begin{tabular}{llcccc} 
\hline \hline \noalign{\smallskip}
Ion  & \hspace{1.5mm}$\lambda$ & $K$     & $\phi_0$ & $\gamma$ & Line \\
     &   (\AA)   &  (\kms) &          &  (\kms)  &       \\[0.5ex]
\noalign{\smallskip} \hline
\noalign{\medskip}
FeI     &  \hspace{-2mm}4957.58 & $343\pm 8$  & 0.014(5) & $-60\pm4$ & em \\ 
KI     & \hspace{-2mm}7664.90  & $351\pm 5$  & 0.021(3) & $-65\pm4$ & em \\
KI      &  \hspace{-2mm}7698.96  & $358\pm 5$  & 0.021(3) & $-61\pm4$ & em \\
FeI     &  \hspace{-2mm}8327.05 & $345\pm 5$  & 0.019(3) & $-58\pm5$ & em \\ 
CaII    &  \hspace{-2mm}8498.02  & $357\pm 3$  & 0.020(4) & $-57\pm5$ & em \\
%TiO edge&  \hspace{-2mm}7589.0   & \hspace{1.5mm}$421\pm 18$ & 0.018(3) & $-63\pm6$ & abs \\ 
NaI     &  \hspace{-2mm}8183.26/8194.82  & $460\pm 5$  & 0.021(2) & $-68\pm3$ & abs \\
%        &  \hspace{-2mm}8194.824  &&&&\\
\noalign{\smallskip} \hline      
\end{tabular}
\end{table}
% -----------------------------------------------------------------------------

Unfortunately, there are no accurate previous measures of
$\gamma$ from optical lines to compare our result with, except the
rather uncertain values quoted by \citet{hellieretal87}.
It is potentially important, however, to note the more positive
$\gamma$ velocities derived from UV and X-ray emission lines that
originate near the white dwarf surface, $\gamma=9.5\pm3.0$\,\kms\
\citep{belleetal05} and $\gamma=-2.8\pm2.3$\,\kms\
\citep{hoogerwerfetal04}. The errors quoted by these authors are
statistical ones and additional systematic errors may affect the
numbers. Nevertheless, it is interesting to note that the difference
between the optical and X-ray/UV results is close to the gravitational
redshift expected for the white dwarf in EX Hya (see
Sect.~\ref{sec:disc}).

%An accurate $\gamma$ value for lines originating from the white dwarf
%offers the possibility to obtain an independent estimate of the mass
%of the white dwarf by measuring the gravitational redshift.
%        1 FeI          4957.597 -55.22
%        2 MgII         4939.396 -62.21
%        6 FeI          4891.492 -63.74
%        9 CrI          4583.901 -58.34
%       13 FeI          4427.299 -63.09
%       21 FeI          4315.084 -57.60
%       25 NeI          4274.660 -58.56
%       26 FeI          4271.759 -57.60
%       29 OI           4233.274 -59.20
%       31 AlII         4226.891 -57.66
%       32 NeII         4226.920 -59.70
%       33 NaI          4216.057 -60.24
%       34 FeI          4198.636 -62.93
%       35 OII          4143.739 -59.81
%       43 ArI          4045.965 -57.29
%       44 MnI          4034.483 -59.59
%       51 CaII         3933.663 -60.53
%       54 FeI          3888.822 -61.60
%       60 FeI          3856.372 -58.08
%       63 HI10         3797.900 -57.78
%       64 HI11         3770.630 -60.10

\subsection{Check on the origin of the narrow emission lines}
\label{sec:test}

\citet{belleetal03} and \citet{hoogerwerfetal04} have reported
accurate values of the radial velocity amplitude of the white dwarf,
$K_1=59.6\pm 2.6$\,\kms\ and $K_1=58.2\pm 3.7$\,\kms, respectively.
We show here that the fully resolved profiles of the narrow optical
lines in our spectra independently yield a very similar, although not
quite as accurate value of $K_1$. To this end, note that the
measurement of two velocity amplitudes that can be assigned to two
points on the line connecting the two stars is equivalent to the
measurement of the radial velocity amplitudes of both stars: we choose
the back of the Roche lobe and the L$_1$ point. The peak velocity in
the NaI absorption line profile at quadrature corresponds to the
velocity amplitude of the back of the star if we account for some
widening by pressure broadening. We choose a point two pixels
(20\,\kms) down into the absorption line profile and obtain
$K_\mathrm{back}=550\pm10$\,\kms. The minimum velocity in the CaII
emission line profiles at quadrature is taken to represent the L$_1$
point. Since the emission lines display no additional broadening, we
choose the first pixel in the CaII line profiles with a non-vanishing
flux and obtain $K_\mathrm{L1}=280\pm10$\,\kms. The ratio of these
velocities equals
$K_\mathrm{back}/K_\mathrm{L1}=x_\mathrm{back}/x_\mathrm{L1}=1.96\pm0.08$,
with $x_\mathrm{back}$ and $x_\mathrm{L1}$ the coordinates along the
$x$-axis of the two extremal points on the stellar surface as measured
from the center of gravity in units of the binary separation. Roche
geometry then implies~$q=M_2/M_1=0.134\pm0.019$,
$x_\mathrm{back}=1.129\pm 0.005$,
$K_2=K_\mathrm{back}/x_\mathrm{back}/(1+q)=430\pm 9$\,\kms,
$K_1=qK_2=58\pm 8$\,\kms, and $K_1+K_2=488\pm 9$\,\kms. With the
orbital period \porb =5895.4\,s and the inclination $i=77.8^\circ$,
the binary separation is $a=(K_1+K_2)P_\mathrm{orb}/(2\pi
\mathrm{sin}i)=(4.683\pm 0.088)10^{10}$cm. Finally, Kepler's law gives
a total mass of $M=0.88\pm0.05$\,\msun\ and component masses
$M_1=0.775\pm 0.045$\,\msun\ and $M_2=0.104\pm 0.016$\,\msun. Since
$K_\mathrm{back}$ and $K_\mathrm{L1}$ were simply read from the
observed profiles, these numbers are approximate only. The important
point to note is the excellent agreement between our value of $K_1$
and the directly measured, more accurate radial velocity amplitude of
the white dwarf quoted at the top of this paragraph. This agreement
verifies our assumption that the low-velocity limit in the CaII line
profiles at quadrature measures the motion of matter that is located
on the secondary star near L$_1$. This assumption forms the basis of
the line synthesis procedure described in Sect.~\ref{sec:model},
below.

% -----------------------------------------------------------------------------
% Table 3
\begin{table}[b]
\caption{Barycentric timings of the spin maxima (spin phase zero),
the eclipse center, and the zero crossing of the absorption lines from
the secondary star (orbital phase zero). The O--C values are
given relative to the ephemerides of \citet{helliersproats92} (see
text).}
\label{tab:timings}
\begin{tabular}{l@{\hspace{7mm}}c@{\hspace{7mm}}c@{\hspace{7mm}}c} 
\hline \hline \noalign{\smallskip}
Band & HJD+2400000 & Error & O--C \\
     &    (d)      &   (d) &  (d)     \\[0.5ex]
\noalign{\smallskip} \hline
\noalign{\medskip}
\multicolumn{2}{l}{\textit{(a) Spin maxima}} \\[0.5ex]
Blue light & 53027.7670 & 0.0014 & 0.0113  \\
               & 53027.8109 & 0.0014 & 0.0087  \\
               & 53030.8381 & 0.0014 & 0.0104  \\
\hbeta\ flux   & 53027.7640 & 0.0014 & 0.0083  \\
               & 53027.8109 & 0.0014 & 0.0087  \\
               & 53030.8360 & 0.0014 & 0.0083  \\[0.5ex]
\multicolumn{2}{l}{\textit{(b) Eclipse timing}} \\[0.5ex]
Blue light, \hbeta\ flux   & 53030.7909 & 0.0014 & \hspace{-1.8mm}$-0.0004$  \\[0.5ex]
\multicolumn{2}{l}{\textit{(c) Absorption line zero crossing}} \\[0.5ex]
NaI doublet    & 53030.7893 & 0.0001 & \hspace{-1.8mm}$-0.0012$  \\			         
\noalign{\smallskip} \hline      
\end{tabular}
\end{table}
% -----------------------------------------------------------------------------

\subsection{Phase shifts}
\label{sec:ephemeris}

All photometrically determined eclipse timings reported over the last
decade occur near a \citet{helliersproats92} orbital phase
$\phi_{98,HS} \simeq 0.98$. The blue-to-red zero crossing of the NaI
line in our data takes place at $\phi_{98,HS}=0.982\pm0.001$
(Tab.~\ref{tab:timings}) indicating that eclipse and zero crossing
coincide within about the jitter in the eclipse timings
\citep[e.g.][]{siegeletal89}. This is confirmed by one pronounced dip
that is superimposed on the first spin maximum in the night of January
26, 2004, occurs at spectroscopic phase zero within the error, and
probably represents the eclipse by the secondary star
(Tab.~\ref{tab:timings}).

The spin light curves in Fig.~\ref{fig:lc} show that blue light, red
light, and the \hbeta\ flux display the same spin modulation within
errors. The U', B and 7500\AA\ bands are integrals over 3760--4000\AA,
4000--4800\AA, and 7450--7550\AA, respectively. The \hbeta\ flux is an
average over $\pm40$\AA\ of the line center corrected for the
underlying continuum. The solid triangles on the abscissa indicate the
times of the spin maxima predicted by the quadratic ephemeris of
\citet{helliersproats92}. The offset between observed and calculated
times of maxima is $\Delta \phi_{67}=-0.20\pm0.03$ and, in what
follows, we use the phase conventions
\begin{equation}
\phi_{98} = \phi_{98,HS}+0.018 \qquad \mathrm{and} \qquad \phi_{67} = \phi_{67,HS}-0.20.
\label{eq:deltaphi}
\end{equation}
Irradiation of the secondary star in EX Hya varies periodically at the
beat period between orbital and rotational periods (210 min) and
inclusion of Eq.~(\ref{eq:deltaphi}) into our illumination model
assures correct phasing.
The only other post-1991 timings of blue light maxima that
we are aware of are those by \citet{eisenbartetal02}, $\Delta
\phi_{67}=-0.04\pm0.01$ at HJD\,2450508, and \citet{belleetal05},
$\Delta \phi_{67}=-0.12\pm0.02$ near HJD\,2451685. Although substantial
fluctuations occur in the timings
\citep{helliersproats92,belleetal05}, the numbers suggest that the
offset increases with time and that the spin-up of the white dwarf in
EX Hya is slightly slower than quoted by \citet{helliersproats92}.
A more regular monitoring of EX Hya with a small telescope a
worthwhile undertaking.

\section{Irradiation model}
\label{sec:model}

We now embark on the construction of a detailed line synthesis model
for NaI$\lambda8183/8195$ and CaII$\lambda 8498$. Our model represents
the Roche lobe by a grid of triangular surface elements, each of which
is characterized by photospheric absorption and potentially by
superimposed emission from a thin layer that we do not distinguish
geometrically from the photosphere.

\subsection{Line profiles}

Velocity smearing is adopted as the dominant broadening mechanism,
supplemented by pressure broadening for the NaI absorption lines and
minimal Gaussian broadening of the CaII emission line. Pressure
broadening is represented by a Lorentz profile derived from the NaI
doublet in the M4.25 dwarf Gl300. The emission lines, on the other
hand, are produced in layers above the photosphere and probably lack
significant pressure broadening. The adopted Gaussian broadening with a
FWHM of one pixel (10\,\kms) merely serves to smooth
irregularities arising from the 0.01 phase bins of the model spectra.

In principle, modeling the complex line profiles that contain
absorption and emission components requires appropriate radiative
transfer calculations for irradiated M-dwarf atmospheres
\citep{brettsmith93,barmanetal04}. Results that could easily be
implemented are not yet available, however, and we consider two simple
cases instead: (i) the fill-up of the absorption line with emission by
adding the two \mbox{components}; and (ii) the gradual disappearance
of the absorption line in the illuminated part and its replacement by
emission. The difference lies in the absorption line wings that are
retained in case (i) and practically disappear in case (ii).  Testing
both models led to better fits with and a clear preference for case
(i) (see Fig.~\ref{fig:f14}, top).  The adopted procedure is
adequate for the present data, but may have to be replaced by a more
sophisticated approach if data of better statistical accuracy become
available.

\subsection{Parameterization of the model}
\label{sec:parameters}

Our kinematic and illumination model has 29 parameters of which 24 are
listed in Tab.~\ref{tab:parameters} and the remaining five are
normalization constants explained in Sect.~\ref{sec:cana}. Parameters
that are kept fixed are the orbital period \porb, the inclination $i$,
the phase shift $\Delta\phi_{67}$, the 210-min amplitude of the
irradiation flux $A_\mathrm{irr}$, the optical depth of the emitting
layer on the secondary $\tau$, the intrinsic FWHM of the line
emission, the NaI gravity darkening coefficient $y_\lambda$, and the
constants $c$ and $d$ in the NaI limb darkening law (all explained
either above or in the next two Sections). We opted to provide the
white dwarf radial velocity amplitude $K_1\simeq 59\pm3$\,\kms\
\citep{belleetal03,hoogerwerfetal04} as a fixed input parameter and
determine its influence on the errors of the fit parameters by
varying it between 56 and 62\,\kms. The mass ratio $q$ is then a
derived parameter and is determined as $q(M,K_1)$. Of the free
parameters, the quantities $M$, $\gamma$, and
$\Delta\phi_{98}$ describe system properties, whereas the disk half
thickness $\beta$, the limit of illumination at $c_\mathrm{lim}$, the
limb darkening/brightening coefficient of the CaII emission
$u_\mathrm{l,Ca}$, the eight numerically given CaII emissivities (or
one parameter of the analytical model), and the five normalization
constants define the emission line model and are discussed in
Sect.~\ref{sec:cana}. For the numerical or analytical versions of the
emission model, 19 or 12 parameters, respectively, are fitted.

% -----------------------------------------------------------------------------
% Table 4
\begin{table}[tb]
\caption{Parameters of the irradiation model for EX Hya and results of
the combined least-squares fit to the NaI$\lambda8183/8195$ and
the CaII$\lambda8498$ line profiles in 16 orbital phase bins.}
\label{tab:parameters}
\begin{tabular}{l@{\hspace{-1mm}}l@{\hspace{1mm}}l@{\hspace{3mm}}ll} 
\hline \hline \noalign{\smallskip}
No. &     Parameter             & \hspace{-3mm}Symbol & free\,/ & Value\,$\pm$\,Error \\
         &            &        & fixed           &            \\[0.5ex]
\noalign{\smallskip} \hline
\noalign{\medskip}
\multicolumn{5}{l}{\textit{(a) System parameters:}} \\[0.5ex]
P1   & Orbital period (s)    &   \porb                 &  fixed   &  5895.4           \\ 
P2   & Inclination           &   $i$                   &  fixed   &  $77.8\pm0.4^{\circ~1)}$              \\ 
P3   & Velocity amplitude (\kms) & $K_1$                &  fixed    &  $59.0\pm3.0^{~2)}$\\
P4   & Total mass (\msun)    &   $M$                   &  free    &  0.898$\pm $0.031    \\ 
P5   & Mass ratio $M_2/M_1$  &   $q$                   &      &  0.137$\pm 0.007^{~3)}$   \\ 
P6   & System velocity (\kms)& $\gamma$                &  free    &  \hspace{-2.0mm}$-58.2\pm1.0$\\
P7   & Phase shift           &$\Delta\phi_{98}$        &  free    & $0.018\pm0.001$\\ 
P8   & Phase shift           &$\Delta\phi_{67}$        &  fixed    &   \hspace{-2.0mm}$-0.20\pm0.06$\\
P9  & Ampl. of irradiation flux & $A_\mathrm{irr}$ &  fixed   &  0.15      \\ [0.5ex]
\multicolumn{5}{l}{\textit{(b) NaI$\lambda8183/8195$  absorption line parameters:}}\\[0.5ex]
P10   & Gravity darkening     & $y_\lambda$             &  fixed   &  0.61          \\ 
P11   & Limb darkening        & $c, d$          &  fixed   &  see text         \\[0.5ex] 
\multicolumn{5}{l}{\textit{(c) CaII and NaI emission line parameters:}} \\[0.5ex]
P12   & Disk half opening angle& $\beta$                 &  free    &  $2.1^{+1.0}_{-0.5}$     \\ 
P13  & Terminator, cos$\,\vartheta$& $c_\mathrm{lim}$            &  free    &  \hspace{-2.4mm}$-0.06\pm0.06$ \\ 
P14  & Optical depth         & $\tau$                   &  fixed   &  $> 10$         \\ 
P15  & Limb darkening        & $u_\mathrm{\,l,Ca}$       &  free   &  $0.42\pm0.24^{~4)}$        \\ 
P16  & FWHM  (\kms)          & $FW$                   &  fixed   &  10         \\[0.5ex] 
P17  & Normalized CaII       & $f_\mathrm{m}(\vartheta)$&  free   &  see Fig.~\ref{fig:eps}      \\ 
\ldots P24 & \hspace{3mm} emission line fluxes&&&\\[0.5ex] 
\noalign{\smallskip} \hline      
\end{tabular}
$^{1)}$ From FWHM of the X-ray eclipse for given WD radius
$R_1(M_1)$.\\ $^{2)}$ From \citep{hoogerwerfetal04,belleetal03}, see text.\\
$^{3)}$ If $K_1$ is fixed in the fit, $q(M,K_1)$ is not an independent
fit parameter.\\ $^{4)}$ Error reflects the strong correlation between $u_\mathrm{\,l,Ca}$ and $M$.
\end{table}
% -----------------------------------------------------------------------------

\subsection{NaI absorption}
\label{sec:na}

\begin{figure}[t]
\begin{center}
\includegraphics[width=8cm]{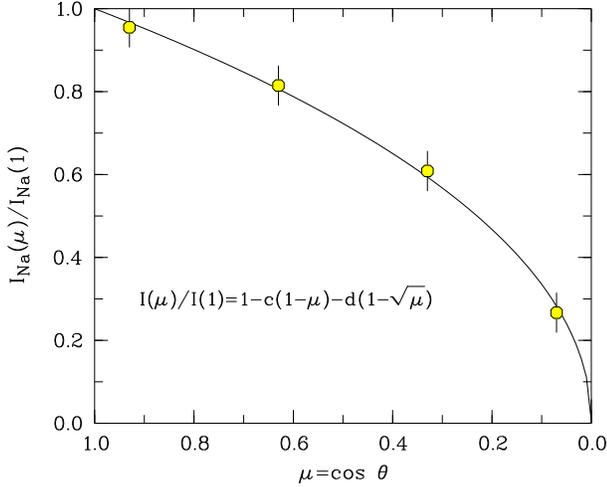}
\caption{Limb darkening of the angle-dependent intensity deficit
integrated over the NaI doublet for a \logg=5 star with \teff= 3000\,K
(Derek Homeier, private communication). The solid line is a fitted
square root law with fit parameters $c$ and $d$ given in the text. }
\label{fig:limb}
\end{center}
\end{figure}

Limb darkening coefficients are available in the literature only for
broad photometric bands. Here, we need the limb darkening of the
integrated NaI absorption line flux. We extract this information from
the model atmosphere of an unirradiated \logg=5 star of 3000\,K kindly
calculated by Derek Homeier with the PHOENIX code. We determine the
mean NaI intensity deficit from the angle-dependent intensity
$I_\lambda(\mu)$ by integration over the doublet as
$I_\mathrm{Na}(\mu)=\int_{line}\left(I_\mathrm{c}(\mu)-I_\lambda(\mu
)\right)d\lambda=EW(\mu)\,I_\mathrm{c}(\mu)$, where
$\mu=\mathrm{cos}\,\theta$ with $\theta$ the zenith angle and
$I_\mathrm{c}$ denotes the continuum outside the line.
Figure~\ref{fig:limb} shows this quantity normalized to the center of
the stellar disk along \mbox{with a square root fit} 
\begin{equation}
I_\mathrm{Na}(\mu)/I_\mathrm{Na}(1)=1-c\,(1-\mu)-d\,(1-\sqrt{\mu}),
\label{eq:limb}
\end{equation}
\citep{claret98} with parameters $c=-0.069$ and $d=1.063$. The EW of
the NaI doublet decreases slightly as the limb is approached and
varies approximately as $\mathrm{EW}(\mu)=(6.72+1.07\mu$)\,\AA. The EW
averaged over the stellar disk is 7.4\AA\ and both, the $EW$ and the
effective temperature of 3000\,K, are typical of a dM5.5 star, the
best estimate of the spectral type of the secondary star in EX Hya
(see Sect.~\ref{sec:disc} and Tab.~\ref{tab:secondary}).

% Fig. 9

We account for monochromatic gravity darkening at 8200\AA\ by a linear
coefficient $y_\lambda=0.61$ (Tab.~\ref{tab:parameters}) that is based
on a surface brightness vs. effective temperature relation derived
from the results of \citet{beuermann06} for field M dwarfs.

Photospheric absorption in the NaI doublet contributed by the $k$-th
surface element at orbital phase $\phi_\mathrm{98,i}$ is then given by
the flux increment 
\begin{equation}
dF_\mathrm{a,k} = -\,a_\mathrm{a}\,dA_\mathrm{k}\,\mu_\mathrm{k}\,
\frac{I_\mathrm{Na}(\mu_\mathrm{k})}{I_\mathrm{Na}(1)}\left(1+2y_\lambda(1-\frac{r_\mathrm{k}}{R})\right),
\label{eq:da}
\end{equation}
 with $dA_\mathrm{k}$ the surface area of the element,
$\mu_\mathrm{k}=\mathrm{cos}\theta_\mathrm{k}$ and $\theta_\mathrm{k}$
the angle between the normal to the element and the line of sight at
orbital phase $\phi_\mathrm{98,i}$, $r_\mathrm{k}$ the radial
separation of that element from the center of the star, and $R$ the
mean stellar radius. The flux increment is collected into the
appropriate 0.3\AA\ wavelength bin that corresponds to the radial
velocity of the element as seen by the observer at
$\phi_\mathrm{98,i}$. For a hidden element with $\mu_\mathrm{k}<0$,
the flux increment is set to zero.  The flux level of the NaI doublet
is determined by the fit parameter $a_\mathrm{a}$ and a further
parameter $a_\mathrm{ratio}$ describes the ratio of the $\lambda8183$
vs. the $\lambda8195$ lines fluxes.

\subsection{CaII and NaI emission}
\label{sec:cana}

Parameterization of line emission from the irradiated atmosphere is
more involved. We model the CaII$\lambda8498$ emission and adopt its
properties for the emission of the NaI doublet with a scaling factor
for the different emission line intensities.  We assume that the
angle-integrated CaII emission line flux $F_\mathrm{Ca}$ of surface
element depends on its distance $r_\mathrm{s}$ from the irradiating
source and on the angle of incidence $\vartheta$ as
$F_\mathrm{Ca}\propto f(\eta)/r_\mathrm{s}^2$ with
$\eta=\mathrm{cos}\,\vartheta$. In the special case of an emitted flux
proportional to the incident energy flux, $f(\eta)\propto \eta$. Model
calculations for $f(\eta)$ in M-dwarf atmospheres are not yet
available, but the case of Lyman line emission of irradiated white
dwarf atmospheres \citep{koenigetal06} suggests that $f(\eta)/\eta$
increases rapidly as grazing incidence is approached ($\eta=0$). We
have tested analytical formulations of $f(\eta)$ and a numerical
tomographic approach. For the latter, we fix $f(\eta)$ at $f_1=0$ for
$\eta=0$ (terminator) and at $f_{10}=1$ for $\eta=0.9$ (L$_1$
point). In between, we define $f(\eta)$ by eight free parameters
$f_{\,2}$ to $f_{\,9}$ at equidistant abscissa values
$\eta^\mathrm{(m)}$.  We interpolate between the $f_\mathrm{m}$ and
regularize the distribution of the emissivities by adding
$A\sum_{m=1}^{9}(f_\mathrm{m+1}-f_\mathrm{m})^2$ to the $\chi^2$ to be
minimized, with $A$ a Lagrange multiplier. This choice implies that a
smoothed version of the current set of $f_\mathrm{m}$ is used as the
default.

The observed light curve of the wavelength-integrated CaII line flux
(Fig.~\ref{fig:calc}, upper panel) is slightly skewed with a centroid
at orbital phase $\phi_{98}=0.517$. We model this asymmetry by
adopting the hot spot or bulge as a second source of irradiation that
we locate in the orbital plane at binary coordinates $x=0, y=0.3$ in a
system with its origin at the center of gravity, unity separation of
both stars along the $x$-axis.

Three further parameters are needed to account for (i) the shadow cast
on the secondary by the accretion disk, (ii) a possible horizontal
energy transfer across the terminator into the dark side of the star,
and (iii) the periodic variation of the irradiation flux received by
the secondary from the rotating magnetic white dwarf. We collect these
dependencies into three additional parameters: (i) the shadow has a
half opening angle $\beta$ independent of azimuth and a sharp edge;
(ii) line emission is allowed to extend beyond the geometric
terminator for illumination by a point source at $\eta=0$ to
$\eta=c_\mathrm{lim}$, where $c_\mathrm{lim}$ is a small negative (or
positive) quantity; and (iii) rotational modulation is modeled by a
variation of the irradiation flux of the form $F_\mathrm{irr} \propto
1+a_\mathrm{irr}\,\mathrm{cos}\,(2\pi\,\phi_\mathrm{irr})$, where
$\phi_\mathrm{irr}=\phi_{67}-\phi_{98}$ is the irradiation phase,
maximum irradiation occurs at $\phi_\mathrm{irr}=0$, and
$F_\mathrm{irr}$ varies with the beat period of 210 min between
orbital and rotational periods. Alas, for the present data, the fit is
insensitive to $a_\mathrm{irr}$, because all exposures cluster around
$\phi_\mathrm{irr}=0.25$ or 0.75. We fix $a_\mathrm{irr}$ at 0.15,
close to the pulsed fractions of blue light \citep{hellieretal87} and
X-ray emission \mbox{\citep{rosenetal91}}.

Finally, we consider the emission properties of the irradiated
atmosphere. The intensity emerging from an infinite isothermal layer
with optical depth $\tau$ along its normal is \mbox{$I \propto
1-\mathrm{exp}(-\tau/\mu )$} with $\mu=\mathrm{cos}\,\theta$. For
large $\tau$, this form approaches the blackbody law and a realistic
model should account for limb darkening or brightening. We find that
all fits with free $\tau$ yield $\tau\ga 10$. Hence, we consider only
the optically thick case that we describe by a linear intensity law
\mbox{$I \propto 1-u_\mathrm{\,l,Ca}(1-\mu)$}, where
$u_\mathrm{\,l,Ca}$ is the limb darkening coefficient for the CaII
line that may be positive or negative. The contribution of surface
element $k$ to the line flux then is
\begin{equation}
dF_\mathrm{e,k} =
a_\mathrm{e,1}\,dA_\mathrm{k}\frac{f(\eta_\mathrm{k})}{r_\mathrm{s}^2}\,
\mu_\mathrm{k} \left(1-u_\mathrm{\,l,Ca}(1-\mu_\mathrm{k})\right),
\label{eq:de}
\end{equation}
where $a_\mathrm{e,Ca,1}$ is the proportionality factor that refers to
the white dwarf as the irradiation source and the flux contribution is
again set to zero for hidden elements
($\mu_\mathrm{k}=\mathrm{cos}\,\theta_\mathrm{k}<0$). Correspondingly,
$a_\mathrm{e,Ca,2}$ describes the 'hot spot' as irradiation source.
The parameter $c_\mathrm{lim}$ that shifts the limit of CaII emission
relative to the geometric terminator of each source is included by
replacing $\eta_\mathrm{k}=\mathrm{cos}\,\vartheta_\mathrm{k}$ in
Eq.~(\ref{eq:de}) by
$\eta'_\mathrm{k}=(\eta_\mathrm{k}-c_\mathrm{lim})/(0.9-c_\mathrm{lim})$
that stays at unity for $\eta=0.9$ (L$_1$ point) and vanishes at
$\eta=c_\mathrm{lim}$. We assume that the NaI emission shares all
parameters with the CaII emission except for its different
normalization relative to CaII expressed by a factor
$a_\mathrm{e,Na}$. The fit yields the parameters listed in
Tab.~\ref{tab:parameters} and, in addition, the values of the five
proportionality factors $a_\mathrm{a}, a_\mathrm{ratio},
a_\mathrm{e,Ca,1}, a_\mathrm{e,Ca,2}$, and $a_\mathrm{e,Na}$.

% Fig. 10
\begin{figure}[t]
\includegraphics[width=7.95cm]{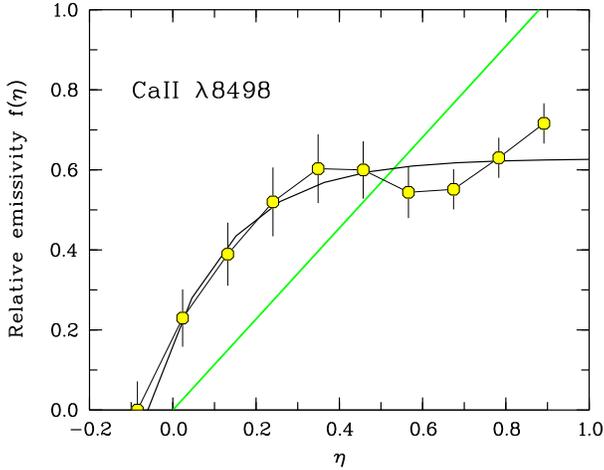}
\caption{Variation of the flux in the CaII$\lambda 8498$ emissivity as
a function of $\eta = \mathrm{cos}\,\vartheta$ for irradiation by a
point source. The green line is for emission proportional to the
energy influx, the black curve, for a simple analytical model that
concentrates the emission closer to the terminator, and the open
circles for the numerical tomographic model.  }
\label{fig:eps}
\end{figure}

\section{Model fits}
\label{sec:fits}

Narrow emission lines have been detected in many CVs and have been
used to determine binary parameters assuming an origin from the
irradiated face of the secondary. We confirm this origin for EX Hya,
but the interpretation of these lines is by no means
straightforward. The centroid of the emission depends on the model
parameters and minimum \chisq\ may occur for different values of the
total mass $M$ depending on the choice of $f(\eta)$ and
$c_\mathrm{lim}$. The situation differs if absorption and emission
lines are considered together because then the model covers the entire
range of radial velocities that occur over the surface of the
secondary star. To obtain an internally consistent fit, we impose the
side condition that the fit assumes minimum $\chi^2$ for the CaII
lines \emph{and} the NaI lines at the \emph{same} $M$, by
appropriately adjusting the parameters of the emission line model. All
fit results presented in this paper comply with this condition.

% Fig. 11
\begin{figure}[t]
\vspace*{0.6mm} \includegraphics[width=8.8cm]{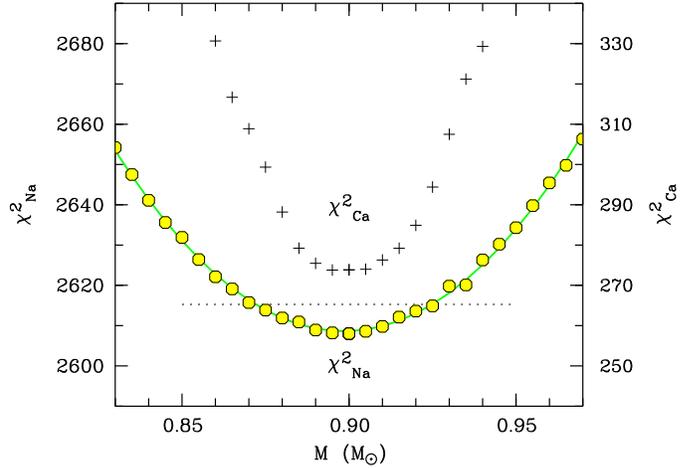}
\caption{Result of the multiparameter fit for the numerical model of
the CaII emissivities of Fig.~\ref{fig:eps}. The NaI value \china\
(open circles) is shown as a function of total mass $M$ with the other
parameters free, the CaII value \chica\ (+) is slaved to reach its
minimum at the same value of $M$ (requiring
$c_\mathrm{lim}=-0.085$). The green curve is a parabola fit and
the dotted line indicates the 99\% confidence level. }
\label{fig:parabola}
\end{figure}

\subsection{CaII$\lambda8498$ emissivity}

A representative set of emission models is shown in
Fig.~\ref{fig:eps}.  The simple assumption of emission proportional to
the incident energy flux, $f(\eta) \propto \eta$ (green line), utterly
fails yielding profiles at quadrature that lack intensity at the
higher velocities and require relatively more emission at small
$\eta$. A one-parameter model of the form $f(\eta) \propto
1-\mathrm{exp}(-\alpha \eta)$ fairs much better. The best fit requires
$\alpha=5.9$ and $c_\mathrm{lim}=-0.060$ (black solid line).  The
tomographic model with eight fitted emissivities $f_\mathrm{m}$
(Sect.~\ref{sec:model}) yields a slightly improved fit. The displayed
model (open circles) is for an intermediate level of regularization
with a Lagrange multiplier $A$\,=\,100 and requires
$c_\mathrm{lim}=-0.085$. Lower and higher values of $A$ produce a more
or less pronounced emission peak near $\eta\simeq0.3$ and the error
bars on the data points reflect the range $A$\,=\,0 to 1000. Despite
the slightly negative values of $c_\mathrm{lim}$ for both models, the
emission is practically limited to the illuminated part of the star if
we consider that the finite extent of the source at the white dwarf
\citep{siegeletal89} alone accounts for a fuzziness of the terminator
of about $2^\circ$ ($\Delta c_\mathrm{lim}=\pm0.035$). Furthermore,
the models lack the resolution to account for an abrupt drop of the
emissivity at $\eta = 0$.  Hence, the slight extension of the emission
beyond the geometric terminator is only marginally significant and we
conclude that there is no compelling evidence for lateral energy
transport in the atmosphere across the terminator. On the other hand,
it is highly significant that the emission decreases much slower with
decreasing $\eta$ than the incident energy flux. The observed increase
in $f(\eta)/\eta$ as the terminator is approached is qualitatively
expected from the calculations of \citet{koenigetal06} for the
hydrogen Lyman line emission from an irradiated (white dwarf)
atmosphere. Realistic calculations are complex and need to consider
the spectral energy distribution of the incident radiation and its
degradation as the terminator is approached in a spherically extended
irradiated atmosphere treated in \mbox{2-D} or \mbox{3-D}.

% Fig. 12
\begin{figure*}[t]
\vspace{1.5mm}
\hspace{9.3mm}
%\hfill
\includegraphics[width=35.0mm,clip]{9010f12a.ps}
\hspace{0.3mm}
\includegraphics[width=35.0mm,clip]{9010f12b.ps}
\hspace{0.3mm}
\includegraphics[width=35.0mm,clip]{9010f12c.ps}
\hspace{1.0mm}
\includegraphics[width=18.10mm,clip]{9010f12d.ps}
\hspace{0.3mm}
\includegraphics[width=18.10mm,clip]{9010f12e.ps}
\hspace{0.35mm}
\includegraphics[width=18.10mm,clip]{9010f12f.ps}

\vspace{-46.2mm}
\includegraphics[width=45.8mm]{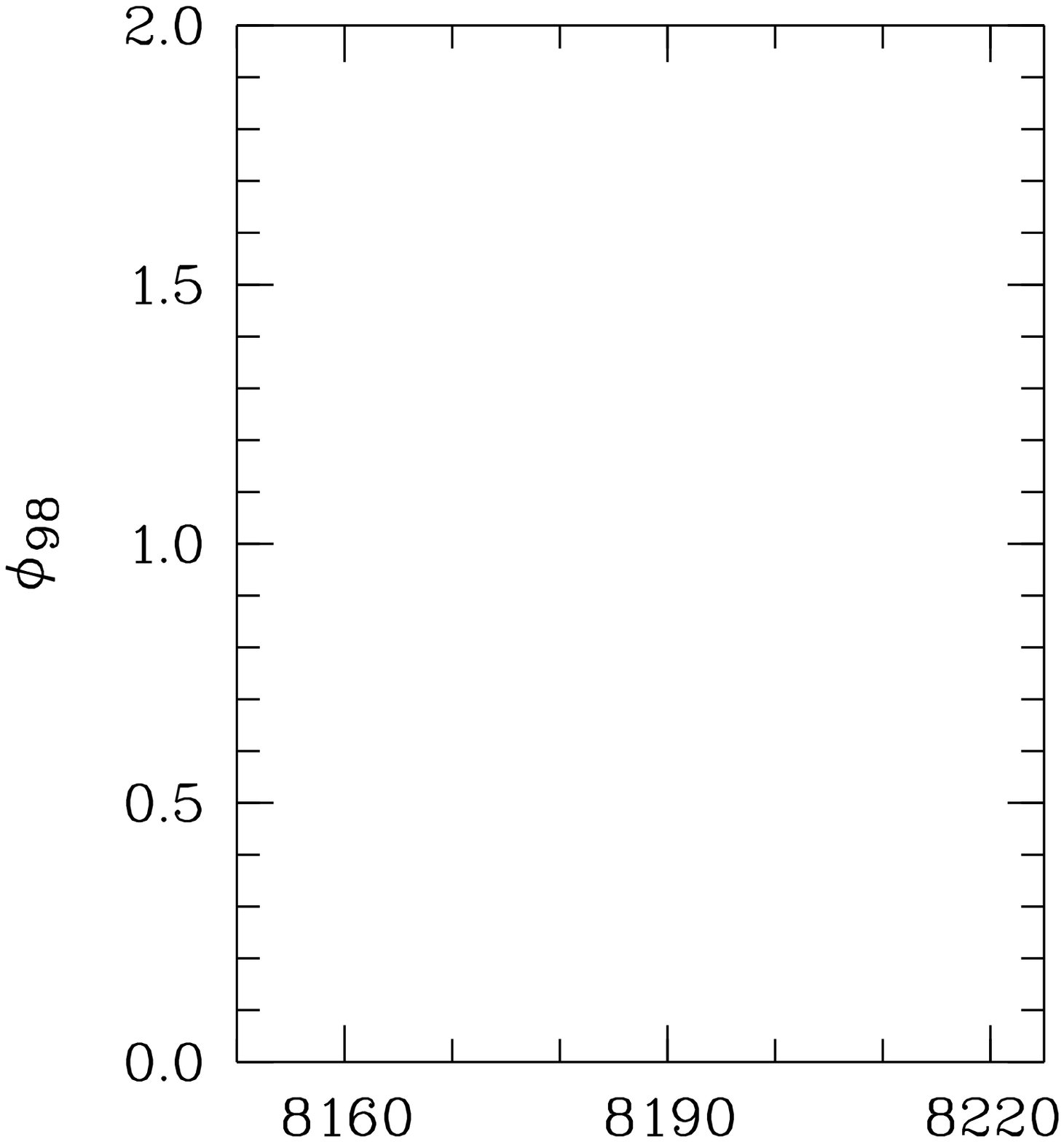}
\hspace{-0.6mm}
\includegraphics[width=35.9mm]{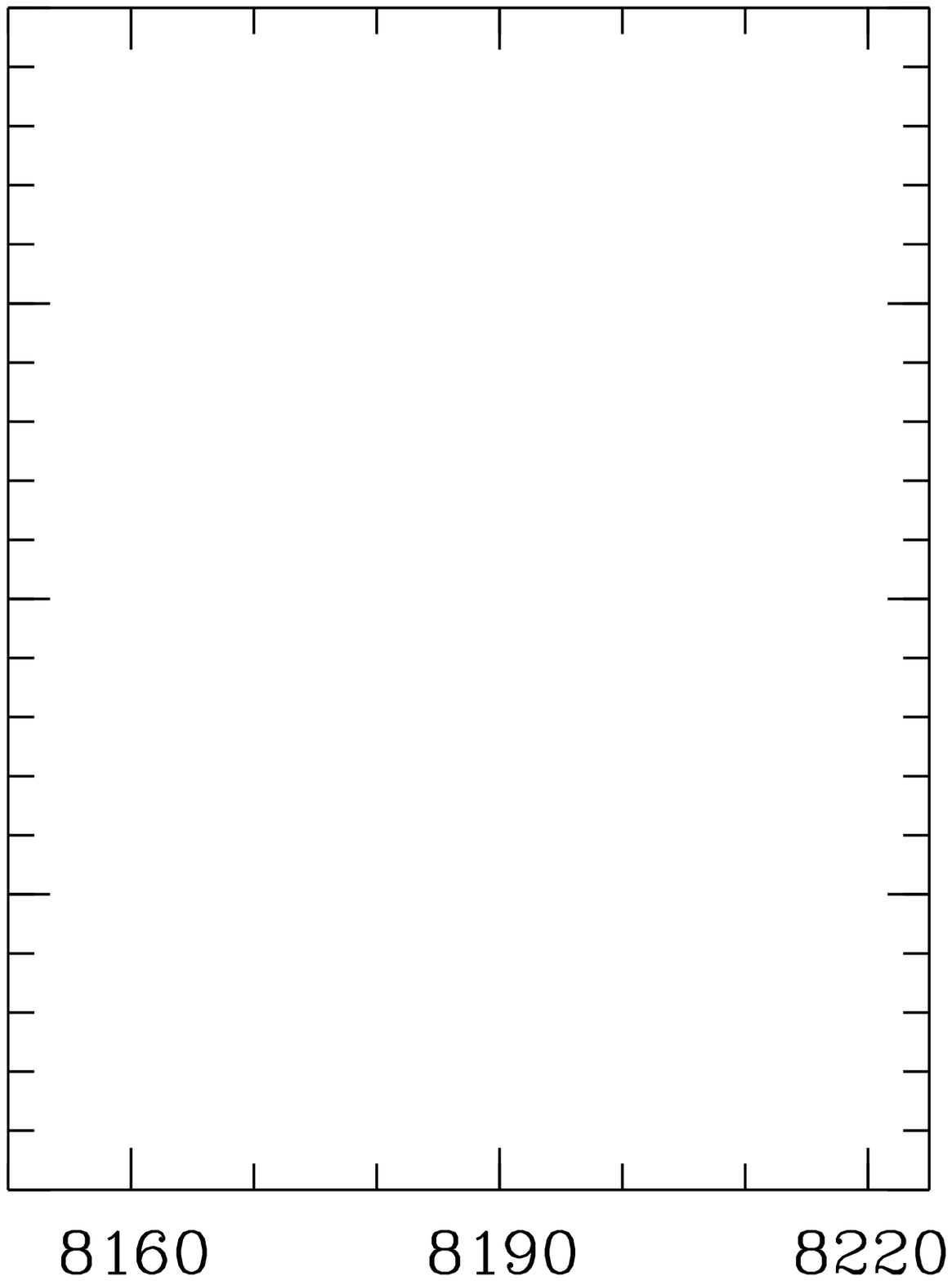}
\hspace{-0.6mm}
\includegraphics[width=35.9mm]{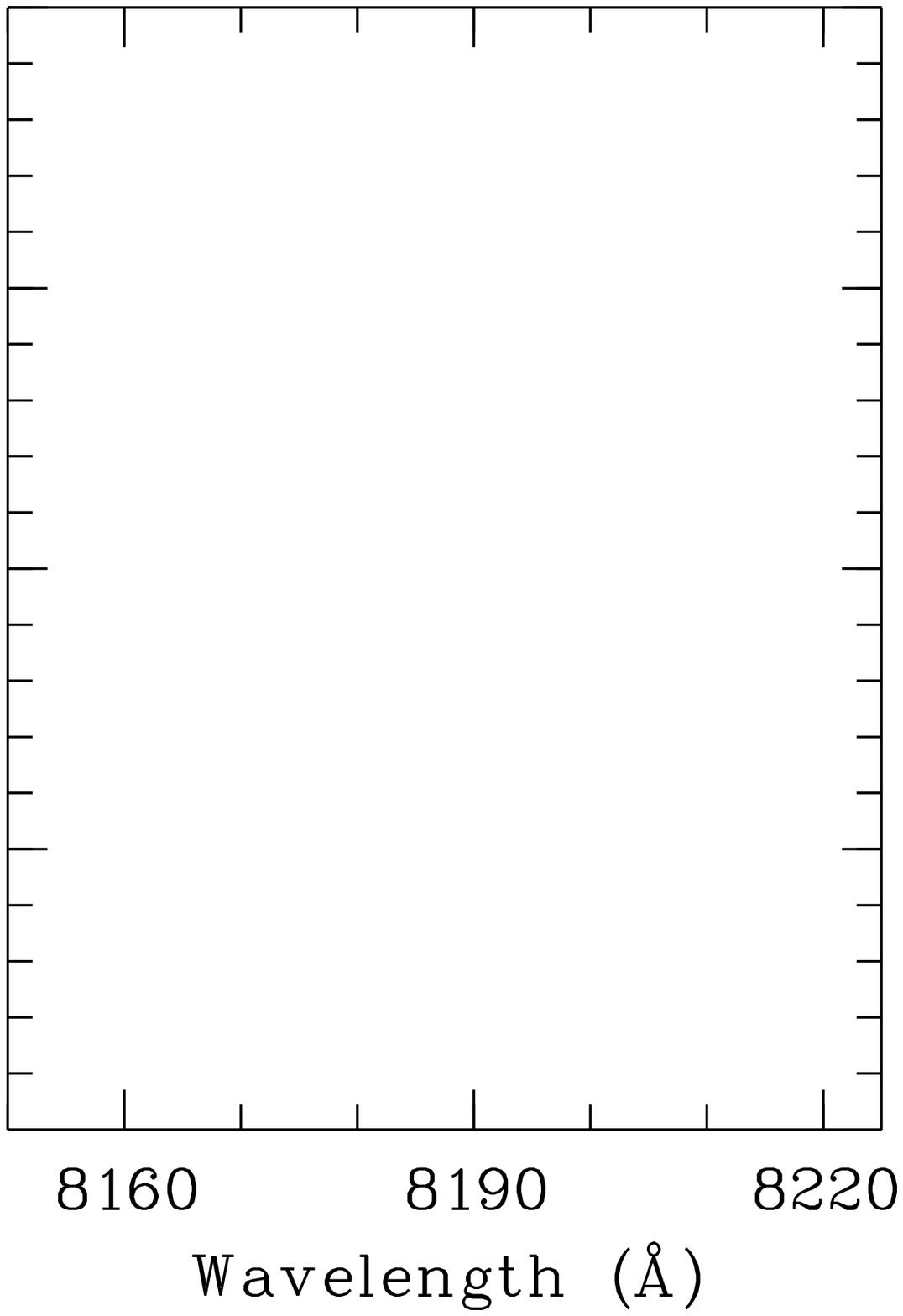}
\hspace{-0.0mm}
\includegraphics[width=18.5mm]{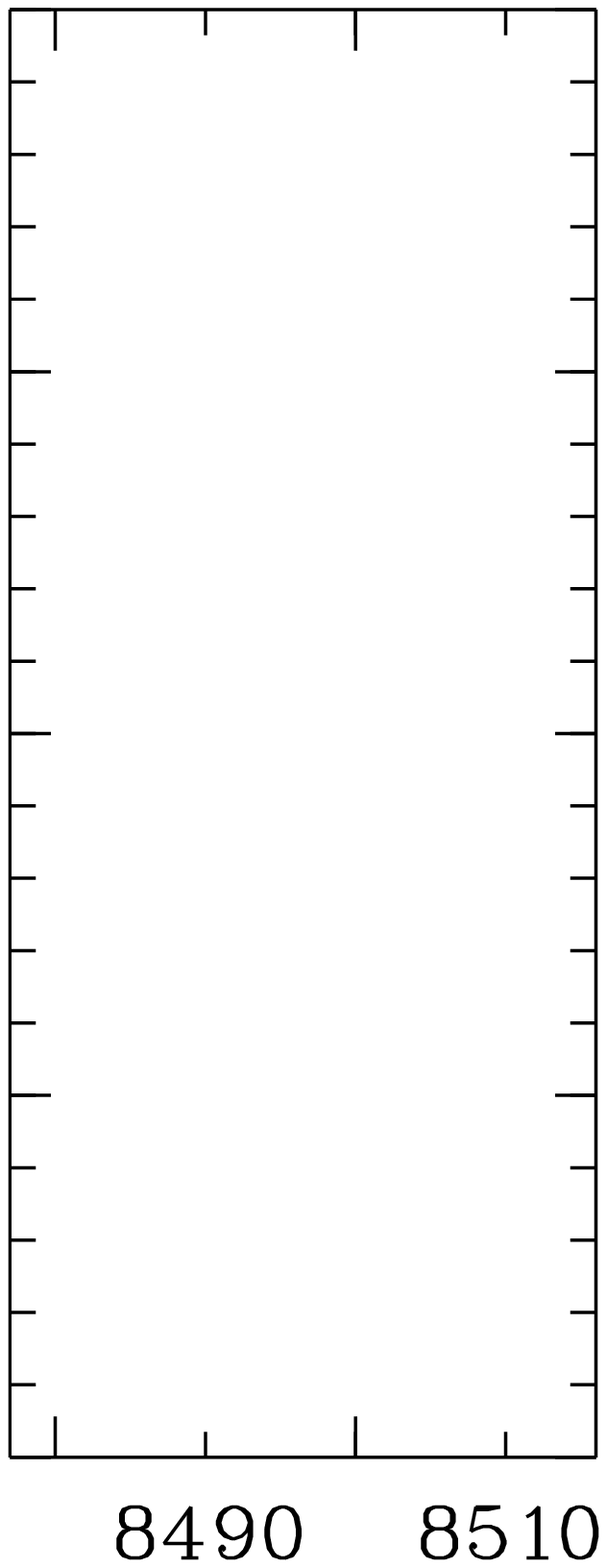}
\hspace{-0.1mm}
\includegraphics[width=18.5mm]{9010r12d.ps}
\hspace{-0.1mm}
\includegraphics[width=18.5mm]{9010r12d.ps}
\caption{{\bf a--f.}~~Sections of Fig.~\ref{fig:phased} containing the
NaI$\lambda8183/8195$ doublet and the CaII$\lambda 8498$ emission
line. The Paschen background has been subtracted in the latter. The
three panels each (a--c and d--f) show the data (left, a, d), the model
(center, b, e), and the residual fluxes (right, c, f). The displayed
intensities are relative to the orbital mean, the zero level is shaded
gray, lower and higher intensities appear darker and lighter,
respectively. FeI$\lambda8220.4$ appears as very faint emission line
in panels a and c, rows 7 to 10. The gray scales in the NaI and CaII
images differ.}
\label{fig:fits1}
\end{figure*}

% Fig. 13
\begin{figure*}[t]
\includegraphics[width=90mm,clip]{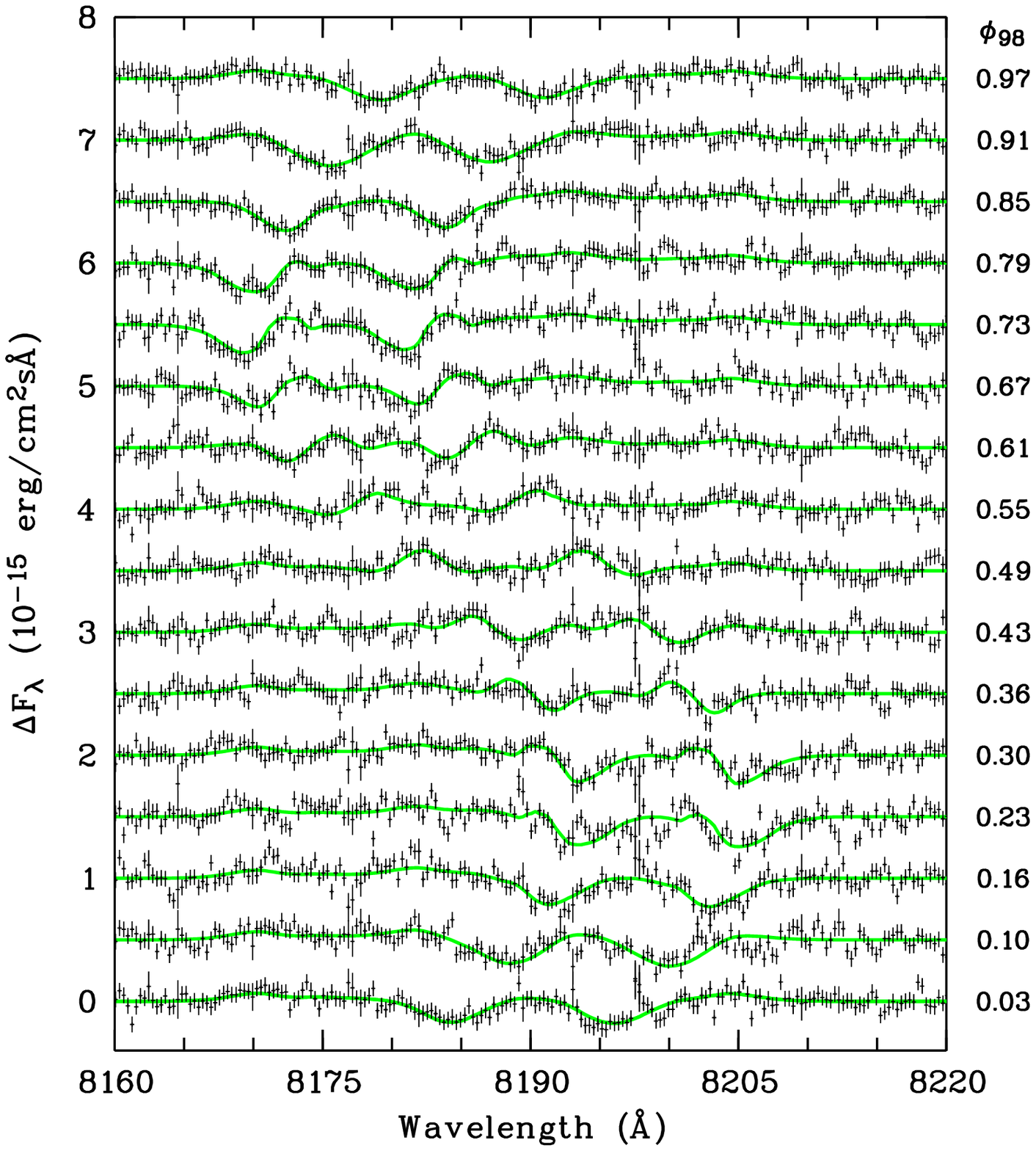}
\hfill
\includegraphics[width=85.5mm,clip]{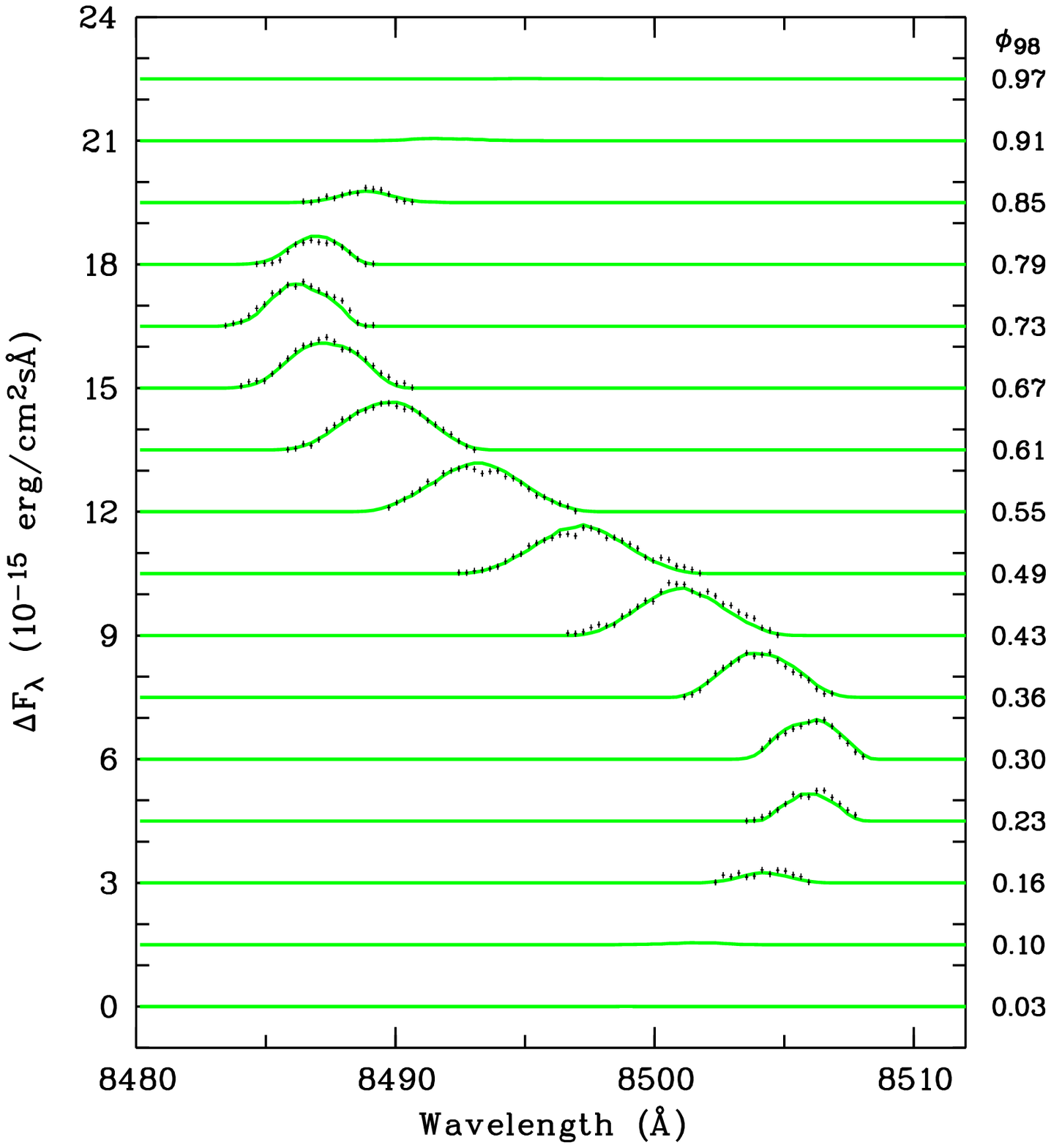}
\caption{{\bf a,b.}~~Comparison of observed phase-resolved line profiles (data
points) and best-fit model spectra (green curves) for the
NaI$\lambda8183/8195$ doublet (left, a) and the CaII$\lambda 8498$ emission
line (right, b). The Paschen background has been subtracted in the latter. }
\label{fig:fits2}
\end{figure*}

% Fig. 14
\begin{figure}[t]
\hspace*{5mm}
\includegraphics[width=7.5cm]{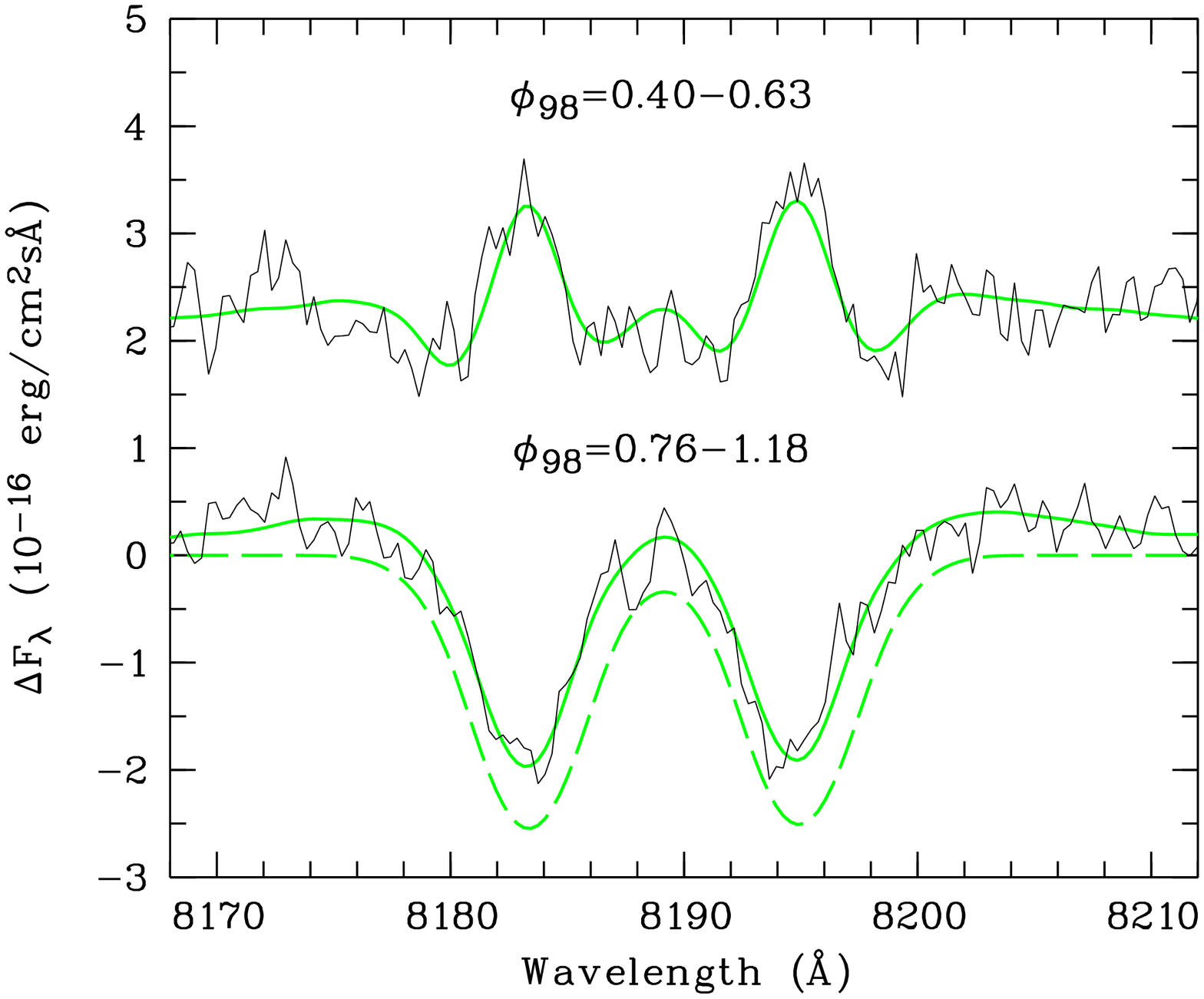}
\caption{Observed (black) and fitted (green) line profiles of the
NaI line for the illuminated (top) and dark side (bottom) of the
secondary star. Same data as in Fig.~\ref{fig:fits2} shifted into the
respective rest system and co-added for two phase intervals (see
Sect.~\ref{sec:cana}). See text for further explanation.}
\label{fig:f14}
\end{figure}

\subsection{Line profiles and quality of the fit}

The high quality of the fit to the phase-resolved line profiles of the
residual spectra (Sect.~\ref{sec:method}) is demonstrated in
Figs.~\ref{fig:fits1} and \ref{fig:fits2}. Figures~\ref{fig:fits1}a and
d show excerpts of Fig.~\ref{fig:phased} that cover the
NaI$\lambda8183/8195$ doublet and the CaII$\lambda8498$ emission line,
respectively, the model line profiles are shown in
Figs.~\ref{fig:fits1}b and e, and the residuals between data and model
in \mbox{Figs.~\ref{fig:fits1}c and f}. Figures~\ref{fig:fits2}a,b display
a quantitative version of the same result. There is no background in the CaII
line profiles since they have been extracted from the profiles of
Fig.~\ref{fig:phased} by removal of the underlying Paschen
background. The enhanced emission at the extremal velocities of the
NaI doublet in Fig.~\ref{fig:fits1}b are a result of the subtraction
of the orbital mean spectrum from the data and the model as described in
Sect.~\ref{sec:method}. The sodium doublet is fitted between 8160.5
and 8214.5\AA\ and contributes 2880 data points in the 16 phase
intervals, while the CaII data set contains 250 non-zero data
points. The best fit has $\chi^2_\mathrm{Na}=2608$ and
$\chi^2_\mathrm{Ca}=274$, or a total~\,$\chi^2=2882$ for 3116
d.o.f. With a reduced \chisqr=\,0.925, the fit is clearly good. The
NaI part benefits from the inclusion of pressure broadening
(Fig.~\ref{fig:fits2}a), while the CaII profiles
(Fig.~\ref{fig:fits2}b) are well matched without any additional
broadening beyond radial velocity smearing. The humps that appear on
the low-velocity slopes of the CaII profiles at \phiorb=\,0.30 and
0.73 are responsible for the enhanced emissivity near $\eta=0.3$ in
Fig.~\ref{fig:eps}. Some variability in the individual CaII profiles
can not be matched by the adjustment of parameters and may indicate
temporal fluctuations of the line emission.

Figure~\ref{fig:f14} shows the spectra observed from the illuminated
and the dark side of the star shifted into their respective rest
systems. The emission line profile (top) displays dips flanking the
emission peaks that represent residues of the incomplete fill-up of
the underlying absorption components. The dips support our choice
of model for the composite NaI line profiles (Sect.~\ref{sec:parameters}).
The observed absorption line profile (bottom) is accompanied by
model spectra \emph{with} and \mbox{\emph{without}} the orbital mean
subtracted (solid and dashed green curves, respectively). They demonstrate
that the subtracted model retains about 90\% of the signal, provided
the spectral structure is restricted to a narrow wavelength range. The
result justifies our use of the mean observed spectrum
as a template for the telluric line correction.

Finally, Fig.~\ref{fig:calc} compares the integrated observed and
modeled CaII emission line fluxes of Fig.~\ref{fig:fits2}b as a
functions of orbital phase (upper panel, black histogram and green
curve). The observed slight asymmetry of the light curve is modeled
by assuming the 'hot spot' as a second source of
irradiation. Some other physical effect can not be excluded.

\subsection{System parameters}
\label{sec:system}

%% Fig. 15
\begin{figure*}[t]
\vspace{2mm}
\hspace*{15.7mm} 
\includegraphics[width=66.3mm,clip]{9010f15a.ps}
\hspace{22.6mm}
\includegraphics[width=66.3mm,clip]{9010f15b.ps}

\vspace{-54.3mm}
\includegraphics[width=88mm]{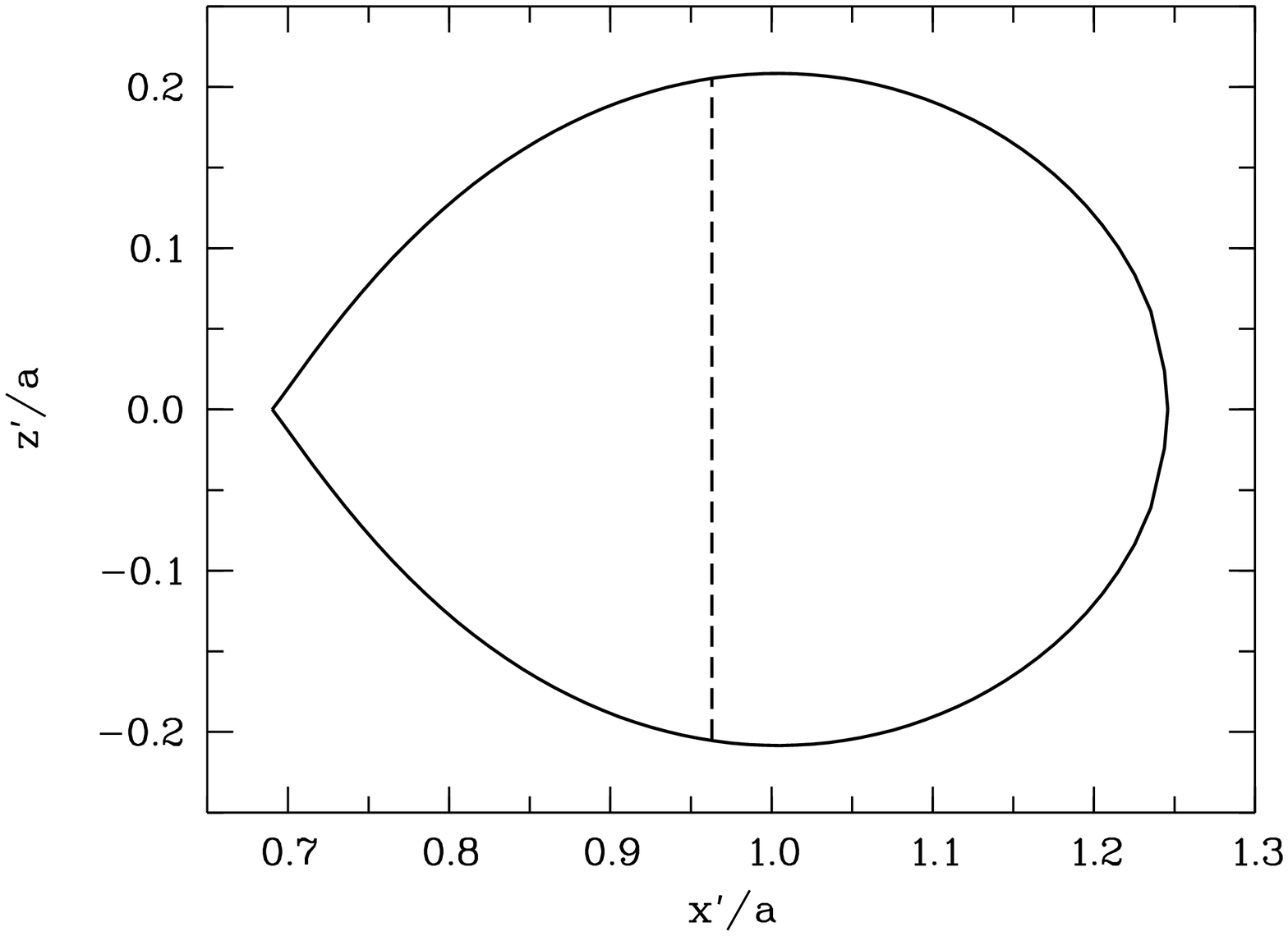}
\hspace{0.8mm}
\includegraphics[width=88mm]{9010r15.ps}
\caption{{\bf a,b.}~~Tomographic images of the secondary star of EX
Hya seen at $\phi_{98}=0.75$ and $i=77.8^\circ$ for the numerical
model of the emissivities of Fig.~\ref{fig:eps}. \emph{(a) Left: }
NaI$\lambda8183/8195$ absorption/emission, \emph{(b) right: }
CaII$\lambda8498$ emission. The $x'-z'$ plane is perpendicular to the
line of sight and coordinates are relative to the binary separation
$a$ with the origin in the white dwarf. The geometric terminator for
illumination by a point source at the white dwarf is indicated by the
dashed line. The intensity scales differ for both images. In the left
hand panel, black is absorption, white emission, and the red rim
represents zero intensity, while the dark side of the star in the
right panel denotes zero line flux that increases over red to
white. Faint emission in the disk shadow and beyond the terminator is
due to irradiation by the 'hot spot'. }
\label{fig:tomo}
\end{figure*}

Figure~\ref{fig:parabola} shows the variation of the \china\ vs. total
mass $M$ for the NaI lines with \chica\ for the CaII emission slaved
to assume a minimum at the same $M$ by appropriate choice of the
parameters of the numerical emission model. There is a well defined
\chisq\ minimum at a total system mass $M=0.898$\,\msun, with an error
of 0.026\,\msun\ at the 99\% confidence level (dotted line). This
result is obtained with a radial velocity amplitude of the white dwarf
of 59\,\kms\ \citep{belleetal03,hoogerwerfetal04} and a system
inclination of $77.8^\circ$. The uncertainties in $K_1$ and $i$ of
about 3\,\kms\ and $0.4^\circ$ add $\pm0.016$\,\msun\ and
$\pm0.005$\,\msun\ to the error budget of $M$, respectively. Adding
the errors quadratically, our measured system mass is $M=0.898\pm
0.031$\,\msun. The remaining fit parameters are summarized in
Tab.~\ref{tab:parameters}.

The slightly mass-dependent inclination $i=(77.8\pm0.4)^\circ$ is
determined from the FWHM of the X-ray eclipse of 155\,s
\citep{mukaietal98,hoogerwerfetal05} on the assumption that only the
lower pole of the white dwarf is eclipsed and is responsible for the
partial character of the X-ray eclipse
\citep{beuermannosborne88,beuermannetal03}. Some variants of the model
\citep{hellieretal87,rosenetal88,rosenetal91} may yield a slightly
different value of $i$ and we note that a $\pm 1^\circ$ change in
$i$ causes a shift of $\mp 0.012$\,\msun\ in $M$.

The velocity amplitude $K_2$ of the secondary star is a derived
quantity and equals $K_2=(2\pi\mathrm{G}M/P)^{1/3}\mathrm{sin}\,i-K_1
= 432.4\pm 4.8$\,\kms. The mass ratio is also a derived quantity and
given by $q=K_1/K_2=0.1365\pm 0.0071$. The error in $K_2$ is almost
entirely due to the noise in the line profiles, while the error in $q$
reflects mostly the uncertainty in $K_1$. The
masses of primary and secondary are $M_1=0.790\pm 0.026$\,\msun\ and
$M_2= 0.108\pm 0.008$\,\msun, respectively. The mean Roche lobe radius
of the secondary star is $R_2=0.1516\pm 0.0034$\,\rsun. Our direct fit
to the line profiles eliminates all problems associated with the
determination of radial velocities from the complex line profiles as
far as possible. Our result for $K_2$ significantly exceeds the value
of $360\pm35$\,kms\ reported by \citet{vandeputteetal03}, on which
most recent published $M_1$ values were based. The system velocity
obtained from the fit is $\gamma=-58.2\pm1.0$\,\kms\ in agreement with
the result obtained in Sect.~\ref{sec:vrad} from the radial velocity
curves.

\subsection{Roche tomography}
\label{sec:tomo}

Roche tomography subjects the emission of all surface elements to some
type of regularization and works best if the binary parameters are
known \citep{watsondhillon04}. Since the determination of these
parameters is our main goal, our tomographic approach involves
the following simplifications: (i) there is no freedom in the contributions of the
individual surface elements to the NaI absorption line profiles
(Sect.~\ref{sec:na}); and (ii) the contributions to the emission line
profiles depend only on the angle of incidence $\vartheta$ of the
irradiation, with shadowing by the accretion disk superimposed
(Sect.~\ref{sec:cana}). However, as discussed above, there is freedom
in the dependence of the emission line flux on the incident energy
flux, in the extent to which the emission extends into the 'dark' side
of the star, and in the angular distribution of the emission line
intensity that emerges from a given surface element
(Sect.~\ref{sec:cana}). The fact that the model fits the data with a
reduced \chisqr=0.92 indicates that any added freedom in the
tomography can not be expected to improve the fit. Such an approach
would need data of much better statistical significance.

Figure~\ref{fig:tomo} shows the images of the secondary star derived
from our restricted tomography as seen at orbital phase 0.75 and an
inclination of $77.8^\circ$. The left panel depicts the NaI image and
the right panel the CaII image. The pictures are based on the
numerical model of the emission line fluxes of Fig.~\ref{fig:eps}.
The principal features are the clear distinction between dark and
irradiated hemispheres of the secondary star and the dark lane
representing the shadow cast by the accretion disk. The geometric
terminator for illumination by a point source at the position of the
white dwarf is indicated by the dashed line. Emission is seen to cease
rather abruptly within a few degrees of the terminator, except for the
faint emission produced by the 'hot spot' that extends beyond the
terminator and fills in the disk shadow. In the NaI image, the effect
of limb (and gravity) darkening is visible as a reduced absorption
line flux as the limb is approached. The shadow of the accretion disk
takes away emission very close to the L$_1$ point, but otherwise has
little effect on the relative distribution of the emission.

\section{Discussion}
\label{sec:disc}

The main result of our high-resolution spectrophotometric study is the
discovery of narrow absorption and emission lines from the photosphere
and chromosphere of the illuminated secondary star in EX Hya. We
present a novel method for the kinematic analysis that involves a
direct fit of a model of the emitting Roche-lobe filling secondary
star to the observed absorption and emission line profiles. A key
quantity is the linear size of the Roche lobe between its back and the
L$_1$-point (Sect.~\ref{sec:test}), which we determine from the
combined fit of the synthesized line profiles to the absorption and
the emission line data. Combined with the previously measured radial
velocity amplitude $K_1$ of the white dwarf
\citep{belleetal03,hoogerwerfetal04}, this line synthesis approach
allows us to determine accurate masses of the binary components.

% -----------------------------------------------------------------------------
% Table 5

\begin{table}[b]
\caption{Parameters of the secondary star in EX Hya compared with
those of Gl551 (Proxima Cen).}
\label{tab:secondary}
\begin{tabular}{lll} 
\hline \hline \noalign{\smallskip}
Parameter &  EX~Hya~B  & Gl551$^{1}$ \\[0.5ex]
\noalign{\smallskip} \hline
\noalign{\medskip}
Spectral type  & dM$5.5\pm 0.5$   & dM5.5             \\
Temperature $T_\mathrm{eff}$ (K)   &                  & $3042\pm117$      \\
Distance $d$ (pc)  & $64.5\pm 1.2$    & $1.295\pm 0.007$  \\
$K$  (mag)     & $12.89\pm0.11$   & 4.36              \\
$M_\mathrm{K}$ (mag) &$8.83\pm 0.12$   & $8.80\pm 0.02$    \\
Mass $M$ (\msun)   & $0.108\pm 0.008$ & $0.123\pm 0.006$  \\
Radius $R$ (\rsun) & $0.1516\pm 0.0034$ & $0.145\pm 0.011$  \\
Velocity amplitude $K_2$ (\kms)   & $432.4\pm 4.8$      &                \\
Systemic velocity $\gamma$ (\kms)   & $-58.2\pm 1.0$      &                 \\[0.5ex]
\noalign{\smallskip} \hline      
\end{tabular}
$^{1)}$ From S{\'e}gransan et al. (2003)\\
\end{table}
% -----------------------------------------------------------------------------

\subsection{The secondary star}

Our fit yields a secondary mass $M_2=0.108\pm0.008$\,\msun\ and a mean
Roche radius $R_2=0.1516$\,\rsun =$1.05\times 10^{10}$\,cm, which
place the star much closer to the main sequence than the former
estimate of 0.078\,\msun\ \citep[e.g.,][]{beuermannetal03}.  The
absolute $K$-band magnitude of EX~Hya~B is $M_\mathrm{K}=8.83\pm0.12$
in perfect agreement with the spectral classification
dM$5.5\pm0.5$. The derived parameters of EX~Hya~B are summarized in
Tab.~\ref{tab:secondary}, with the data of the nearby dM5.5 star Gl551
that has an interferometrically measured radius
\citep{segransanetal03} added for comparison. The radius of an
0.108\,\msun\ model star of solar composition is $0.925\times
10^{10}$\,cm \citep{baraffeetal98} and rises to $0.98\times
10^{10}$\,cm if rotational deformation is accounted for
\citep{renvoizeetal02}. These models apply to stars without spots,
while spotted stars tend to have somewhat larger radii. Just how much
this effect influences the radii of the secondary stars in CVs is not
known (see Beuermann et al. 2006 and references therein). The
calibration of the radii of field stars vs. $M_\mathrm{K}$
\citep[][Eq.~7]{beuermannetal99}~\,yields $R=0.1514$\,\rsun\ for
$M_\mathrm{K}=8.83$, confirming that EX~Hya~B is essentially a main
sequence star, but differs from a field star of the same absolute
magnitude by being \mbox{deformed and irradiated}.

\subsection{The white dwarf}

Previous estimates of the masses of the two stellar components in EX
Hya relied on the uncertain velocity amplitude of the secondary star
of \citet{vandeputteetal03}, $K_2=360\pm35$\,\kms. Our result,
$K_2=432.4\pm4.8$\,\kms, is 2.1$\sigma$ from Vande Putte's result and
yields a much more accurate kinematic solution. The primary mass
$M_1=0.790\pm0.026$\,\msun\ supersedes the value of
$M_1\simeq0.46$\,\msun\ quoted in most recent studies on EX Hya
\citep[e.g.][]{beuermannetal03,vandeputteetal03,
hoogerwerfetal04,hoogerwerfetal05,mhlahloetal06}~\footnote{The primary
mass of 0.91\msun\ derived by \citet{belleetal03} from $K_1=59.6$
\kms, $K_2=360$ \kms, $i=78^\circ$, and an assumed $M_2=0.152$\,\msun\
violates Kepler's law.}. The new mass is close to the mean for
short-period CVs, $\langle M\rangle\simeq 0.8$\,\msun, and confirms
the early result $M_1=0.78\pm0.17$\,\msun\ of
\citet{hellieretal87}. Given the larger primary mass, one can
re-estimate the mass transfer rate following the analysis of
\citet[][see also Ritter (1985)]{beuermannetal03}. From their Fig.~1,
$\dot{M_1}$ is seen to drop with increasing $M_1$ and reach about the
value expected from gravitational radiation as the sole momentum
transfer process for $M_1=0.79$\,\msun.

The radius of a white dwarf of $0.790\pm 0.026$\,\msun\ with an
intrinsic effective temperature of about 15000\,K is $R_1=(7.35\pm
0.23)\times 10^8$\,cm based on models with a thick hydrogen envelope
\citep{wood95}. The mean temperature of the white dwarf in EX~Hya
determined from the HST FOS spectrum is 25000\,K
\citep{eisenbartetal02}, but that temperature is dominated by the
heated polar caps responsible for the pronounced spin modulation seen
in X-rays and in the UV and the temperature of the underlying white
dwarf is likely to be lower, roughly as noted above. With the quoted
radius, the gravitational redshift at the surface of the white dwarf
is $\upsilon_\mathrm{grav}=47.6\pm3.1$\,\kms. The predicted apparent
systemic velocity of the white dwarf then is
$\gamma+\upsilon_\mathrm{grav}=-10.6\pm 4.1$\,\kms. The velocity
derived from the X-ray emission lines that originate close to the white
dwarf surface is $\gamma=-2.8\pm2.3$\,\kms\ \citep{hoogerwerfetal04},
where the error is the statistical one and the systematic error is
larger (C. Mauche, private communication). While the interpretation of
these numbers in terms of a gravitational redshift measurement may be
premature, it is clear that an independent measurement of the white
dwarf mass becomes feasible with a more secure value of the apparent
systemic velocity of the white dwarf.
 
\begin{acknowledgements}
We thank Ansgar Reiners for providing the Gl406 spectrum, Derek
Homeier for his result on the limb darkening of the NaI line flux,
Peter Hauschildt for his model atmosphere results, and all of them for
helpful discussions. Joachim Krautter and Nikolaus Vogt acquired the
unpublished UBVRIJHK photometry of EX Hya. We also thank our
colleagues Christopher W. Mauche, Frederic V. Hessman, Hans Ritter,
and Axel Schwope for comments on an earlier draft of this paper.
\end{acknowledgements}

\bibliographystyle{aa}

\end{document}